\documentclass[fleqn,usenatbib]{mnras}

\usepackage{newtxtext,newtxmath}

\usepackage[T1]{fontenc}

\DeclareRobustCommand{\VAN}[3]{#2}
\let\VANthebibliography\thebibliography
\def\thebibliography{\DeclareRobustCommand{\VAN}[3]{##3}\VANthebibliography}

\usepackage{graphicx}	
\usepackage{amsmath}	
\usepackage{comment}

\title[Forming Uranus and Neptune concurrently]{Can Uranus and Neptune form concurrently via pebble, gas and planetesimal accretion?}  

\author[L.E.J. Eriksson et al.]{
Linn E.J. Eriksson,$^{1}$\thanks{E-mail: linn.eriksson@stonybrook.edu}
Marit A.S. Mol Lous,$^{2, 3}$
Sho Shibata$^{2}$ and 
Ravit Helled$^{2}$
\\
$^{1}$ Institute for Advanced Computational Sciences, Stony Brook University, Stony Brook, NY, 11794-5250, USA \\
$^{2}$ Institute for Computational Science, University of Z\"urich, Winterthurerstr. 190, CH-8057 Zurich, Switzerland \\
$^{3}$ Weltraumforschung und Planetologie, Physikalisches Institut, Universit\"at Bern, Gesellschaftsstrasse 6, 3012 Bern, Switzerland
}

\date{Accepted XXX. Received YYY; in original form ZZZ}

\pubyear{2023}

\begin{document}
\label{firstpage}
\pagerange{\pageref{firstpage}--\pageref{lastpage}}
\maketitle

\begin{abstract}
The origin of Uranus and Neptune has long been challenging to explain, due to the large orbital distances from the Sun. After a planetary embryo has been formed, the main accretion processes are likely pebble, gas and planetesimal accretion. Previous studies of Uranus and Neptune formation typically don't consider all three processes; and furthermore, do not investigate how the formation of the outer planet impacts the inner planet. In this paper we study the concurrent formation of Uranus and Neptune via both pebble, gas and planetesimal accretion. We use a dust-evolution model to predict the size and mass flux of pebbles, and derive our own fit for gas accretion. We do not include migration, but consider a wide range of formation locations between 12 and $40\, \textrm{au}$. If the planetary embryos form at the same time and with the same mass, our formation model with an evolving dust population is unable to produce Uranus and Neptune analogues. This is because the mass difference between the planets and the H-He mass fractions become too high. However, if the outer planetary embryo forms earlier and/or more massive than the inner embryo, the two planets do form in a few instances when the disk is metal-rich and dissipates after a few Myr. Furthermore, our study suggests that {\it in-situ} formation is rather unlikely. Nethertheless, giant impacts and/or migration could potentially aid in the formation, and future studies including these processes could bring us one step closer to understanding how Uranus and Neptune formed.
\end{abstract}

\begin{keywords}
planets and satellites: general -- planets and satellites: formation -- planets and satellites: gaseous planets
\end{keywords}



\section{Introduction}
Uranus has a mass of of $14.5\, \textrm{M}_{\oplus}$ and a semimajor axis of $19.1\, \textrm{au}$, placing it around $10\, \textrm{au}$ beyond the orbit of Saturn. The second ice giant Neptune has a mass of $17.1\, \textrm{M}_{\oplus}$ and a semimajor axis of $30.0\, \textrm{au}$, placing it at the inner edge of the Kuiper Belt and around $10\, \textrm{au}$ beyond Uranus' orbit. Unlike the Solar System gas giants, Uranus and Neptune are primarily composed of heavy elements\footnote{all elements heavier than helium} with a hydrogen-helium-fraction $\lesssim 20\%$ (see \citealt{HelledBodenheimer2014} and references therein). 

The formation of Uranus and Neptune has been challenging for the planet formation models since several decades. \citet{Safronov1972} demonstrated that the time-scale for core formation via planetesimal accretion exceeds the lifetime of the disk at the orbits of Uranus and Neptune. This time-scale issue can be alleviated by instead considering the accretion of small mm-cm sized pebbles (e.g. \citealt{Lambrechts2014,VenturiniHelled2017}). However, when the accretion of gas is properly considered, the challenge reverses and the difficulty is to prevent the planets from undergoing runaway gas accretion and becoming gas giants \citep{HelledBodenheimer2014}. This problem can be circumvented by removing the gas disk at the exact right time, which is known as the fine-tuning problem. Furthermore, both pebble accretion and planetesimal accretion have an efficiency which is generally decreasing with orbital distance, making it hard to explain why Neptune has a mass that is higher than Uranus. 

Some of the above concerns can be alleviated if the planets have not formed in-situ. In fact, in order to explain multiple properties of the Kuiper Belt and the Oort cloud, it has been suggested that Neptune must have migrated outwards to reach its current location (e.g. \citealt{Malhotra1995,HahnMalhotra1999,Nesvorny2015}). In \citet{Nesvorny2015}, it was found that Neptune must have been located interior of $25\, \textrm{au}$ and slowly migrated outwards over a timescale longer than $10\, \textrm{Myr}$. A formation location closer to the Sun implies a more efficient core accretion, making it possible to form the planets over a shorter time-scale. In the Nice model, Uranus and Neptune are formed at distances of $\sim 12-20\, \textrm{au}$, before an instability occurs which causes them to move outwards and eventually end up at their current orbital locations \citep{Tsiganis2005}. During this instability, Uranus and Neptune also switched places in some of their simulations, a mechanism which can explain why Neptune has a mass higher than that of Uranus.

The aforementioned studies concerned late migration that occurred after the dispersal of the gas disk; however, planets can also undergo migration during the disk lifetime due to interactions with the surrounding gas disk (see \citealt{Paardekooper2022} for a recent review on planet migration). The direction and magnitude of this migration depends on both planet and disk properties. Interactions with dust in the disk can further modify the migration properties \citep{Guilera2023}. There is no evidence which suggests that Neptune could not have had an early period of migration driven by interactions with the protoplanetary disk, prior to the late outward migration which placed it at its current location. The existence of the classical Kuiper belt beyond $40\, \textrm{au}$ does put some constraints on the initial formation locations, since the planets could not have formed in or passed through that region without disturbing the planetesimal belt. 

Taken together, the formation of Uranus and Neptune is still very unconstrained and suffers numerous challenges. \citet{VallettaHelled2022} used a pebble accretion model coupled with a realistic gas accretion model to study the possibility of forming Uranus and Neptune in-situ, and found that they could indeed obtain planets with the right masses and H-He mass fractions. In this work, we develop this formation model further by considering the accretion of both pebbles, gas, and planetesimals in an evolving disk. We use a state-of-the art dust evolution model and also account for the blocking of pebbles towards Uranus by accretion onto Neptune. Furthermore, we consider a wide range of formation locations and disk parameters. The aim of our study is to investigate whether both planets could form concurrently without invoking additional growth mechanisms such as giant impacts.  


We present our models for disk evolution, planet growth and describe our simulation set-up in Section \ref{sect:model}. The results for in-situ formation are shown in Section \ref{sect:in-situ}, and in Section \ref{sect:vary location} we show the results from when the formation locations are varied. In Section \ref{sect:potential solutions} we discuss potential mechanisms that can change our results compared to the previous sections, and finally our main findings are summarised in Section \ref{sect:conclusions}. 

\section{Our model}\label{sect:model}
We consider growth via pebble, gas, and planetesimal accretion in an evolving 1D global disk model. We use two methods for determining the pebble flux: in the \texttt{constant} model the pebble flux is proportional to the gas flux through the disk, while in the \texttt{evolving} model pebble flux is calculated using a dust evolution code. We derive a fit for gas accretion based on MESA simulations \citep{Paxton_2011, Paxton_2013, Paxton_2015, Paxton_2018, Paxton_2019}. Planetesimal accretion is modelled using a semi-analytic model and assuming that the initial planetesimal surface density is proportional to the gas surface density. The effect of planetary migration is not included; however, we do test a large range of formation locations.  

\subsection{Disk model}
The evolution of the disk's surface density is modelled using the analytic solution for an unperturbed thin accretion disk from \citet{Lynden-Bell+1974}:
\begin{align}
    \Sigma_\mathrm{\rm disk}(t) = \frac{\dot{M}_{\rm disk}}{3 \pi \nu_\mathrm{out} (r/r_\mathrm{out})^\gamma} \exp \left[{-\frac{(r/r_\mathrm{out})^{2-\gamma}}{T_\mathrm{out}}}\right],
\end{align}
where $T_\mathrm{out} = t/t_s +1$ and $t_s = 1/(3 [2-\gamma]^2) \times r_\mathrm{out}^2/\nu_\mathrm{out}$.
In the above expressions $t$ is the time, $\dot{M}_{\rm disk}$ is the disk accretion rate, $\nu_{\rm out}$ is the kinematic viscosity at semimajor axis $r=r_{\rm out}$ and $\gamma$ is radial viscosity gradient. We adopt $r_{\rm out}=50\, \textrm{au}$ and $\gamma=15/14$. The kinematic viscosity is given by:  
\begin{equation}
    \nu = \alpha \Omega_K H_g^2,
\end{equation}
\citep{ShakuraSunyaev1973}, where $\alpha$ is the viscosity parameter, $\Omega_K$ is the Keplerian angular velocity and $H=c_s/\Omega_K$ is the scale height of the gas disk. We use $\alpha=0.005$ throughout this study. The sound speed is calculated as $c_s = (k_B T/[\mu m_H])^{1/2}$ where $K_B$ is the Boltzmann constant, $\mu$ is the mean molecular weight, $m_H$ is the mass of the hydrogen atom and $T=150\, \textrm{K} (r/\textrm{au})^{-3/7}$ is the mid-plane temperature of the disk \citep{ChiangGoldreich1997}. The value of $\mu$ is chosen to be $2.34$.

The evolution of the disk's accretion rate is given by: 
\begin{equation}
    \dot{M}_{\rm disk} (t) = \dot{M}_{0,\rm disk} T_\mathrm{out}^{-(5/2-\gamma)/(2-\gamma))} \left[1 - \left(\frac{t}{t_{\rm disk}}\right)^{3/2}\right],
\end{equation} 
where the last term is implemented to mimic gradual disk dissipation until time $t=t_{\rm disk}$ \citep{Ruden2004}. The radial velocity of the disk gas at semimajor axis $r$ can be obtained from the continuity requirement using the expression:
\begin{equation}
    v_{\rm R,disk} = -\frac{\dot{M}_{\rm disk}}{2\pi r \Sigma_{\rm disk}}.
\end{equation}

\subsection{Pebble accretion model}
Growth via pebble accretion is modelled using the exact monodisperse accretion rate from \citet{Lyra2023}:
\begin{equation}
    \dot{M}_{\rm pe} = \pi R_{\rm acc}^2 \rho_{\rm pe} \delta v \exp{(-\xi)}[I_0(\xi) + I_1(\xi)],
\end{equation}
where 
\begin{equation}
    \xi = \left(\frac{R_{\rm acc}}{2H_{\rm pe}}\right)^2.
\end{equation}
In the above equations $I_X$ are modified Bessel functions of the first kind of real order, $R_{\rm acc}$ is the accretion radius, $\rho_{\rm pe}$ is the density of pebbles in the mid-plane, $\delta v$ is the approach speed and $H_{\rm pe}$ is the pebble scale height.

We that all pebble accretion occur in the Hill regime, and thus the approach speed and accretion radius can be approximated as: 
\begin{equation}
    \delta v = \Omega_K R_{\rm acc}
\end{equation}
and
\begin{equation}
    R_{acc}=({\rm St}/0.1)^{1/3} R_H,
\end{equation}
where $\rm St$ is the Stokes number of the pebbles and $R_{\rm H}$ is the Hill radius of the planet \citep{Johansen2017}. Our assumption of Hill accretion can sometimes overestimate the accretion rate when we consider small ($<0.1-1\, \textrm{M}_{\oplus}$) planetary masses and Stokes numbers, where accretion should occur in the Bondi regime (see e.g., Fig. 5 in \citealt{Lyra2023}). Since the difference in accretion rate is small near the transition from Bondi to Hill accretion, and  the Bondi regime is rather limited in our simulations, this should not have a large impact on our results. Furthermore, \citet{Lyra2023} demonstrated that the pebble accretion rate  changes when considering a distribution of pebble sizes rather than a single size. The polydisperse accretion rate can then be significantly higher than the monodisperse rate in the Bondi regime, and slightly lower in the Hill regime.

The pebble density in the disk's mid-plane is given by:
\begin{equation}
    \rho_{\rm pe} = \frac{\Sigma_{\rm pe}}{\sqrt{2\pi}H_{\rm pe}}.
\end{equation}
We calculate the pebble scale height as $H_{\rm pe} = H \sqrt{\alpha_{\rm T}/(\alpha_{\rm T} + {\rm St})}$ \citep{KlahrHenning1997,LyraLin2013}, where $\alpha_{\rm T}$ is the turbulent parameter (note that the turbulent parameter is not the same as the viscous parameter $\alpha$ that regulates the viscous evolution of the gas disk). The pebble surface density is obtained from the continuity requirement $\Sigma_{\rm pe} = \dot{M}_{\rm pf}/(2\pi r v_{\rm R,pe})$. 

We use two different models for determining the radial pebble flux past the planet location $\dot{M}_{\rm pf}$ and the Stokes number. In the \texttt{constant} model, we adopt a constant Stokes number and $\dot{M}_{\rm pf} = Z \times \dot{M}_{\rm disk}$, where $Z$ is the disk metallicity. These are common assumptions used in the literature; however, as demonstrated by \citet{Drazkowska2021} the growth of planets can change significantly when considering a more realistic dust evolution model. Motivated by this, we also construct an \texttt{evolving} model, where we use pebble fluxes and Stokes numbers from the code \texttt{pebble predictor} \citep{Drazkowska2021}. The \texttt{pebble predictor} is a semi-analytic model which predicts the pebble flux and flux-averaged Stokes number as a function of time and semimajor axis in an arbitrary unperturbed disk. This model is heavily dependent on the assumed fragmentation velocity $v_{\textrm{frag}}$ and radial extent of the disk $R_{\textrm{edge}}$. We consider two different fragmentation velocities, $v_{\textrm{frag}}=1\, \textrm{m}s^{-1}$ and $10\, \textrm{m}s^{-1}$, and three different disk extents, $R_{\textrm{edge}}=50\, \textrm{au}$, $100\, \textrm{au}$ and $200\, \textrm{au}$. Further details on the \texttt{pebble predictor} are presented in  Appendix \ref{sect:pebble predictor}. Furthermore, when calculating the radial pebble flux towards the inner planet, we remove the pebbles accreted onto the outer planet. 

The radial drift velocity of pebbles with Stokes number ${\rm St}<<1$ can be approximated as: 
\begin{equation}
    v_{\rm R,pe} = -2 {\rm St} \eta v_{\rm K} + v_{\rm R,disk},
\end{equation}
where 
\begin{equation}
    \eta = -\frac{1}{2}\left(\frac{H}{r}\right)^2 \frac{\partial \ln P}{\partial \ln r}.
\end{equation}
In the above expression $v_{\rm K}$ is the Keplerian orbital velocity and $P=\Sigma_{\rm disk}T/H$ is the disk's pressure. 

In our simulations we allow the accretion of pebbles to continue until the total planetary mass exceeds the pebble isolation mass ($M_{\rm iso}$), which can be calculated by \citep{Bitsch2018}:
\begin{equation}
    M_{\rm iso} = 25\, \textrm{M}_{\oplus} \times f_{\textrm{fit}},
\end{equation}
where
\begin{equation}
    f_{\textrm{fit}} = \left[ \frac{H/r}{0.05} \right]^3 \left[ 0.34\left( \frac{\log(\alpha_3)}{\log (\alpha_{\rm T})} \right)^4 + 0.66 \right] \left[ 1 - \frac{\partial \ln P / \partial \ln r + 2.5}{6} \right],
\end{equation}
and $\alpha_3=10^{-3}$.

\subsection{Gas accretion model}
The rate of gas accretion of a growing planet is highly uncertain, and there are various different prescriptions presented in the literature. Rather than picking one of these prescriptions, we chose to derive our own fit for gas accretion based on planet formation simulations performed with the MESA code that was updated to model planetary formation \citep{Valletta_2020}. These simulations are carried out using the same disk parameters and pebble-accretion prescription as our \texttt{constant} model. We simulate 30 random cases and self-consistently calculate the gas accretion rate. The resulting gas accretion can be well represented by an analytical fit: 
\begin{align}
    \dot{M}_{\text {gas,1}} &=
        10^{a_{1}}
        \left( \frac{M_{\text {core }}}{1 \textrm{M}_{\oplus}} \right)^{b_{1}}
        \left( \frac{\dot{M}_{\text {solid }}}{10^{-7} \textrm{M}_{\oplus} / \mathrm{yr}} \right)^{c_1} 
        \textrm{M}_{\oplus} / \mathrm{yr}, \label{eq:dMenvdt_1} \\ 
    \dot{M}_{\text {gas,2}} &=
        10^{a_{2}}
        \left( \frac{M_{\text {core }}}{1 \textrm{M}_{\oplus}} \right)^{b_{2}}
        \left( \frac{M_{\text {env }}}{1 \textrm{M}_{\oplus}} \right)^{c_{2}}
        \left( \frac{\dot{M}_{\text {solid }}}{10^{-7} \textrm{M}_{\oplus} / \mathrm{yr}} \right)^{d_2}
        \textrm{M}_{\oplus} / \mathrm{yr}, \label{eq:dMenvdt_2}
\end{align}
and the accretion rate can be obtained as: 
\begin{align}
    \dot{M}_\mathrm{gas} =\dot{M}_\mathrm{gas,1} +\dot{M}_\mathrm{gas,2}.  \label{eq:dMenvdt_fit}
\end{align}
The parameters of the inferred fit are presented in Table \ref{tab:fitting_dMgas}. The fit to the gas accretion depends on the planetary core mass, envelope mass, and solid accretion rate. Although we only considered pebble accretion in the MESA simulations, in our model the solid accretion rate is taken as the sum of pebble accretion and planetesimal accretion. The fitting parameters further depend on the chosen grain opacity. 
We consider two different scaling factors for the contribution of grains in the opacity $f_{\textrm{g}}$: $0.1$ and $1.0$. Further details on the MESA simulations and the construction of the fit are given in Appendix \ref{sect: appendix gas accretion}. 

\begin{table}
	\centering
	\caption{Parameters of the fitting function for gas accretion.}
	\label{tab:fitting_dMgas}
	\begin{tabular}{cccc} 
		\hline
        & $f_{\rm g}=1.0$ & $f_{\rm g}=0.1$ \\
    \hline
    $a_1$ & -8.655389 & -8.058656 \\
    $b_1$ & 3.488167 & 3.262527 \\
    $c_1$ & -0.449784 & -0.464667 \\
    \hline
    $a_2$ & -10.725292 & -11.188670 \\
    $b_2$ & 3.989025 & 5.834267 \\
    $c_2$ & 2.415257 & 2.880980 \\
    $d_2$ & -0.307779 & -1.116815 \\
                    \hline
	\end{tabular}
\end{table}

Similar to \citet{Bitsch2015}, we limit the gas accretion rate onto the planet to 80\% of the disk accretion rate, as it has been shown that not even deep gaps can fully halt the radial flow of gas \citep{LubowDangelo2006}. Since there is no significant gas accretion expected during the earliest phases of core formation, we turn on gas accretion after the core reaches a mass of $1\, \textrm{M}_{\oplus}$. Furthermore, since neither Uranus nor Neptune have massive gas envelopes, we only consider gas accretion in the regime $M_{\rm core}>M_{\rm env}$. Whenever a planet reaches $M_{\rm core}<M_{\rm env}$ in our simulations, we stop the simulation and record the total masses. Finally, we implement a floor value for the solid (i.e. heavy-element)  accretion rate of $\dot{M}_{\rm solid}=10^{-10}\, \textrm{M}_{\oplus}\textrm{yr}^{-1}$, in order to prevent problems with the fit as the gas disk dissipates and $\dot{M}_{\rm solid} \rightarrow 0$.

\subsection{Planetesimal accretion model}
The location and timing of planetesimal formation in the Solar nebula is highly disputed. Population studies often assume a smooth wide-stretched disk of planetesimals, whereas simulations of planetesimal formation tend to promote formation in specific regions of the disk. Examples of such locations are around ice-lines (\citealt{DrazkowskaAlibert2017,SchoonenbergOrmel2017}), at local pressure bumps (e.g. \citealt{Carrera2021}) and at planetary gap-edges (\citealt{Stammler2019,Eriksson2020}). Formation in a wider region of the disk has been shown to be possible in the case of efficient gas removal via photoevaporation \citep{Carrera2017}. In this study, for simplicity, we assume that the initial planetesimal surface density $\Sigma_{\rm pl}$ is equal to the planetesimal metallicity ($Z_{\textrm{pl}}$) times the gas surface density at the start time of the simulation ($t_{\textrm{start}}$). We consider planetesimal metallicities of $0$ (core growth only occurs via pebble accretion), $0.25Z$ and $0.5Z$. The pebble metallicity is then taken to be $Z-Z_{\textrm{pl}}$. The time $t_{\textrm{start}}$ is varied in the parameter study, and since the gas surface density decreases with time, this implies that planets which begin to form late will be surrounded by a planetesimal disk of lower mass than planets which begin to form early.

The planetesimal accretion rate is calculated using the semi-analytic model of \citet{Chambers2006}: 
\begin{equation}
    \dot{M}_{\rm pl} = \frac{2\pi \Sigma_{\rm pl} R_H^2}{P_{\rm orb}} P_{\rm coll}, \label{eq:planetesimal_accretion_rate}
\end{equation}
where $P_{\rm orb}$ is the planet's orbital period and $P_{\rm coll}$ is the mean collision rate. We use the mean collision rates from \citet{Inaba2001}. 
$P_{\rm coll}$ depends on the eccentricity and inclination of the planetesimals, and the capture radius of a protoplanet. The evolution of the eccentricity, inclination,  and surface density of the planetesimal disk are modelled using the statistical approach \citep[e.g.][]{Inaba2001}.
In this study, we use the model developed by \cite{Fortier_2013}. 
This calculation takes into account gas drag as well as viscous stirring from the planet and the surrounding planetesimal disk. The collision radius is approximated using eq. 7 from \citet{VallettaHelled2021} when the total mass is above $2\, \textrm{M}_{\oplus}$ and the H-He mass fraction is above 1\%. For lower masses and H-He mass fractions, the collision radius is set to be the core radius, calculated using a density of $1455\, \textrm{kg}\textrm{m}^{-3}$ (an average between the density of Uranus and Neptune). We consider planetesimals of $100\, \textrm{km}$ in size, and with a bulk density of $1000\, \textrm{kg}\textrm{m}^{-3}$. Further details of the planetesimal accretion model are presented in Appendix \ref{sect:appendix planetesimal accretion}. 

\subsection{Simulation set-up}

\begin{table}
	\centering
	\caption{Values of used parameters in this study.}
	\label{tab:pebb current ps}
	\begin{tabular}{lccccr} 
		\hline
        $\mathbf{St}$ (\texttt{constant}) & 5e-3 & 6e-3 & 7e-3 & 8e-3 & 9e-3 \\
        & 1e-2 & 2e-2 & 3e-2 & 4e-2 & 5e-2 \\
        \hline
        $\mathbf{Z}$ & 5e-3 & 6e-3 & 7e-3 & 8e-3 & 9e-3 \\
        & 1e-2 & 2e-2 & 3e-2 & 4e-2 & 5e-2 \\
        \hline
        $\mathbf{Z_{\textrm{pl}}}$ & 0 & 0.25Z & 0.5Z & & \\
        \hline
        $\mathbf{\alpha}_\mathrm{\mathbf{T}}$ & 1.0e-5 & 2.5e-5 & 5.0e-5 & 7.5e-4 & 1.0e-4 \\
        & 2.5e-4 & 5.0e-4 & 7.5e-4 & 1.0e-3 &  \\
        \hline
        $\dot{\mathbf{M}}_\mathrm{\mathbf{0,disk}}$ & 1e-8 & 2e-8 & 3e-8 & 4e-8 & 5e-8 \\
        $[\mathrm{M}_{\odot}\mathrm{yr}^{-1}]$ & 6e-8 & 7e-8 & 8e-8 & 9e-8 & 1e-7 \\
        \hline
        $\mathbf{t}_{\mathrm{\mathbf{start}}}$ & 1e5 & 1e6 & 2e6 & & \\
        $[\mathrm{yr}]$ & & & & & \\ 
        \hline
        $\mathbf{t}_{\mathrm{\mathbf{disk}}}$ & 3e6 & 5e6 & 10e6 & & \\
        $[\mathrm{yr}]$ & & & & & \\
        \hline
        $\mathbf{f}_{\mathrm{\mathbf{g}}}$ & 0.1 & 1 & & & \\
		\hline
        $\mathbf{R}_{\mathrm{\mathbf{edge}}}$ (\texttt{evolving}) & $50$ & $100$ & $200$ & & \\
        $[\textrm{au}]$ & & & & & \\
        \hline
        $\mathbf{v}_{\mathrm{\mathbf{frag}}}$ (\texttt{evolving}) & 1 & 10 & & & \\
        $[\mathrm{m}\textrm{s}^{-1}]$ & & & & & \\
		\hline
    \end{tabular}
\end{table}

We adopt a linear time-grid with a time-step of $dt=500\, \textrm{yr}$, and use a simple Euler method to update the planetary masses. We initiate our planetary embryos at the same time and with a mass of $0.01\, \textrm{M}_{\oplus}$. 
As we discuss below, we also perform simulations where the outer planet is inserted at a higher mass, to mimic an earlier and/or more massive embryo formation. In our \texttt{constant} model, we vary the Stokes number, disk metallicity, fraction of solid mass in planetesimals versus pebbles, turbulent alpha, initial disk accretion rate, time of embryo formation, disk evaporation time and grain opacity. In our \texttt{evolving} model, we calculate the Stokes number using the \texttt{pebble predictor} code, and instead vary the radial extent of the protoplanetary disk. In the \texttt{evolving} model, we further vary the fragmentation velocity. The values of all parameters that are varied in the parameter study are listed in Table \ref{tab:pebb current ps}. For each configuration we perform $\sim 500,000$ simulations using the \texttt{constant} model, and $\sim 300,000$ simulations when using the \texttt{evolving} model.

In the first part of the paper, we consider in-situ formation where Uranus and Neptune are located at $19.1\, \textrm{au}$ and  $30.0\, \textrm{au}$, respectively. In the second part of the paper, we vary the planetary formation locations, and also consider the option that the planets could have switched places after the dissipation of the gas disk. For simplicity, we assume that the planets do not migrate or shift location during the disk's lifetime. Therefore, when varying the formation locations, we do not consider the potential mechanisms that eventually led to the current semimajor axes of Uranus and Neptune at $19.1\, \textrm{au}$ respectively $30.0\, \textrm{au}$. This could have occurred due to a period of dynamical instability after the dispersal of the gas disk, as suggested by the Nice model.

\section{In-situ formation}\label{sect:in-situ}
\begin{figure}
    \includegraphics[width=\columnwidth]{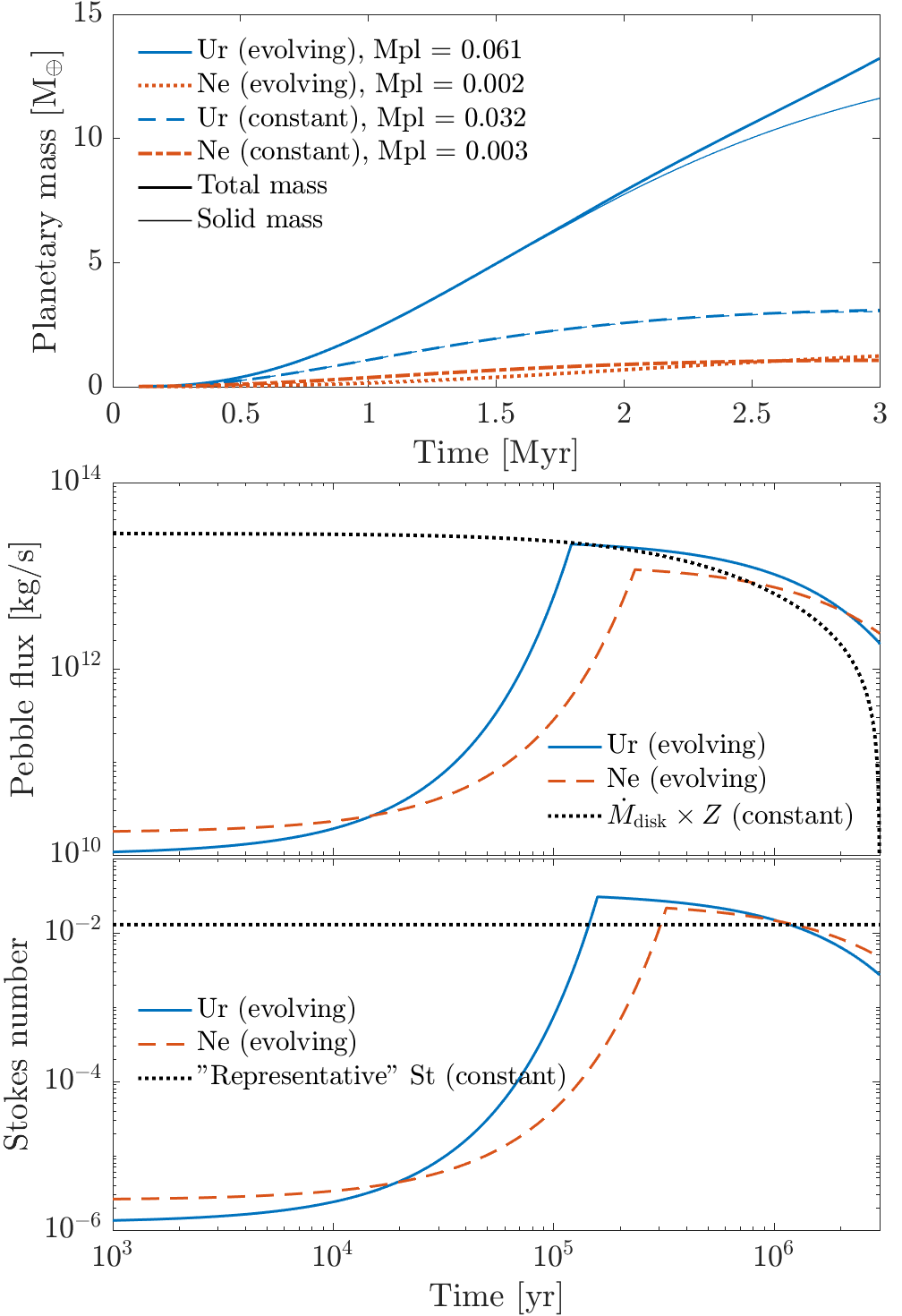}
    \caption{The growth of Uranus and Neptune in-situ.
    The top panel shows the growth-tracks for one example simulation, produced using the parameters given in Table \ref{tab: 1 case parameters}. The \texttt{evolving} model results in a higher mass for Uranus than the \texttt{constant} model, and a larger mass difference between the two planets. The corresponding pebble flux and Stokes number evolution are shown in the two bottom panels. 
    }
    \label{fig:1case}
\end{figure}

We first investigate the possibility of forming both planets at their current semimajor axes. The growth-tracks for one example simulation with the parameters listed in Table. \ref{tab: 1 case parameters} are shown in Fig. \ref{fig:1case}, along with the corresponding time evolution of the pebble flux and Stokes number. For these given parameters, it takes $\sim 10^5\, \textrm{yr}$ for the solid population to grow until maximum size in the \texttt{evolving} model. The growth takes approximately twice as long at Neptune's location compared to Uranus' location,  resulting in a smaller maximum Stokes number and pebble flux. However, the radial drift is slower at Neptune's location, which leads to a slower decline of the pebble flux with time. In this example, the pebble flux resulting from the \texttt{evolving} model declines at a slower rate than it does in the \texttt{constant} model. Furthermore, since the time evolution of the gas disk in neglected in the calculation of the Stokes number and pebble flux in the \texttt{pebble predictor}, the pebble flux from the \texttt{evolving} model does not equal zero at $t=t_{\rm disk}$. 

In order to demonstrate how the growth-tracks change when switching between the \texttt{constant} model and the more realistic \texttt{evolving} model, we calculate a "representative" Stokes number and use that in the \texttt{constant} model. We obtained this "representative" Stokes number as follows. We calculate the time-average of the flux-weighted Stokes number at both planet locations; and take the average of the two numbers. The resulting growth-tracks are presented in the top panel of Fig. \ref{fig:1case}. The \texttt{evolving} model produces a rather good Uranus analogue; however, the corresponding mass of Neptune is just above $1\, \textrm{M}_{\oplus}$. The \texttt{constant} model results in a much lower mass for Uranus, and in general a smaller mass difference between the two planets. The growth-tracks for a second example simulation are shown in Appendix \ref{sect: 2 selected cases}, where we further demonstrate how the planet growth changes when considering: only pebble accretion; pebble and gas accretion; and finally pebble, gas, and planetesimal accretion. 

\begin{table}
	\centering
	\caption{Parameters used to produce the example simulation presented in Fig. \ref{fig:1case}.}
	\label{tab: 1 case parameters}
	\begin{tabular}{ll} 
	\hline
    "Representative" St (\texttt{constant}) & 0.0129 \\
    $Z$ & 0.01 \\
    $Z_{\rm pl}$ & $0.5Z$ \\
    $\alpha_{\textrm{T}}$ & $10^{-5}$ \\
    $\dot{M}_{0,\textrm{disk}}$ & $9\times 10^{-8}\, \textrm{M}_{\odot}\textrm{yr}^{-1}$ \\
    $t_{\textrm{start}}$ & $10^5\, \textrm{yr}$ \\
    $t_{\textrm{disk}}$ & $3 \times 10^6\, \textrm{yr}$ \\
    $f_{\rm g}$ & 1.0 \\
    $R_{\textrm{edge}}$ (\texttt{evolving}) & $200\, \textrm{au}$ \\
    $v_{\rm frag}$ (\texttt{evolving}) & $10\, \textrm{ms}^{-1}$\\
    \hline
	\end{tabular}
\end{table}

\begin{figure*}
\centering
{\includegraphics[width=1.4\columnwidth]{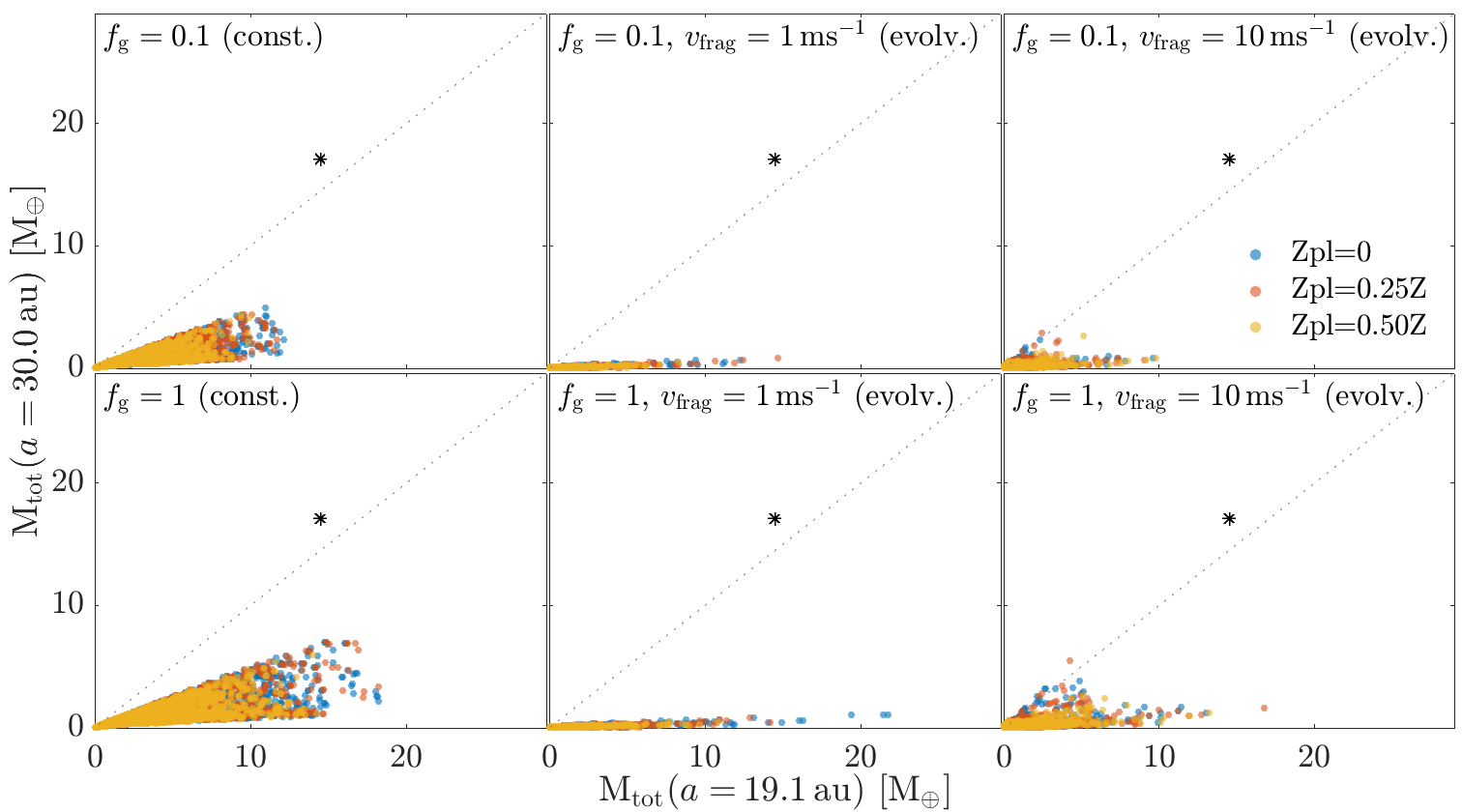}}
\caption{Plots showing the results of all our {\it in-situ} simulations where the H-He mass fraction is less than 20\% at the time of disk dissipation. Simulations in which the H-He mass fraction becomes larger than 20\% are not shown on the plots. The masses of Uranus and Neptune is marked with an asterix. Although we consider a wide range of parameters, no Uranus and Neptune analogues are found.}
\label{fig: scatt pebbGas}
\end{figure*}

In Fig. \ref{fig: scatt pebbGas}, we show the outcome of all the simulations when assuming {\it in-situ} formation. The scatter points indicate the total masses of all the planets with a H-He mass fraction less than 20\% at the time of disk dissipation (we calculate the H-He mass fraction as the ratio of gas mass to total mass). Although we used a wide range of parameters, we do not  produce any Uranus and Neptune analogues. Most models do manage to produce an Uranus analogue, although they are rare, but the corresponding mass of Neptune in these cases is always well below $10\, \textrm{M}_{\oplus}$. This is not unexpected, since both pebble and planetesimal accretion tend to be more efficient at smaller semimajor axes.

When a fragmentation velocity of $1\, \textrm{m}\textrm{s}^{-1}$ is being used (middle column), the dust-evolution model struggle to grow Neptune above $1\, \textrm{M}_{\oplus}$. This problem becomes less severe when the fragmentation velocity is increased, and in this scenario there are a few cases when Neptune grows to become more massive than Uranus. This is either an effect of the dust-evolution model, and/or due to the blocking of pebbles by Neptune. When we use $f_{\rm g}=0.1$, the gas fractions become too high in the mass range relevant to Uranus and Neptune. As the fraction of mass in planetesimals vs pebbles increases, the total masses of planets with H-He mass fractions below 20\% in the \texttt{constant} simulation decreases. This is expected since pebble accretion tends to be more efficient than planetesimal accretion at large semimajor axes. This trend is harder to spot when looking at the results from the \texttt{evolving} model, since the simulation outcome in general is much more variable. 

\begin{figure}
    \includegraphics[width=\columnwidth]{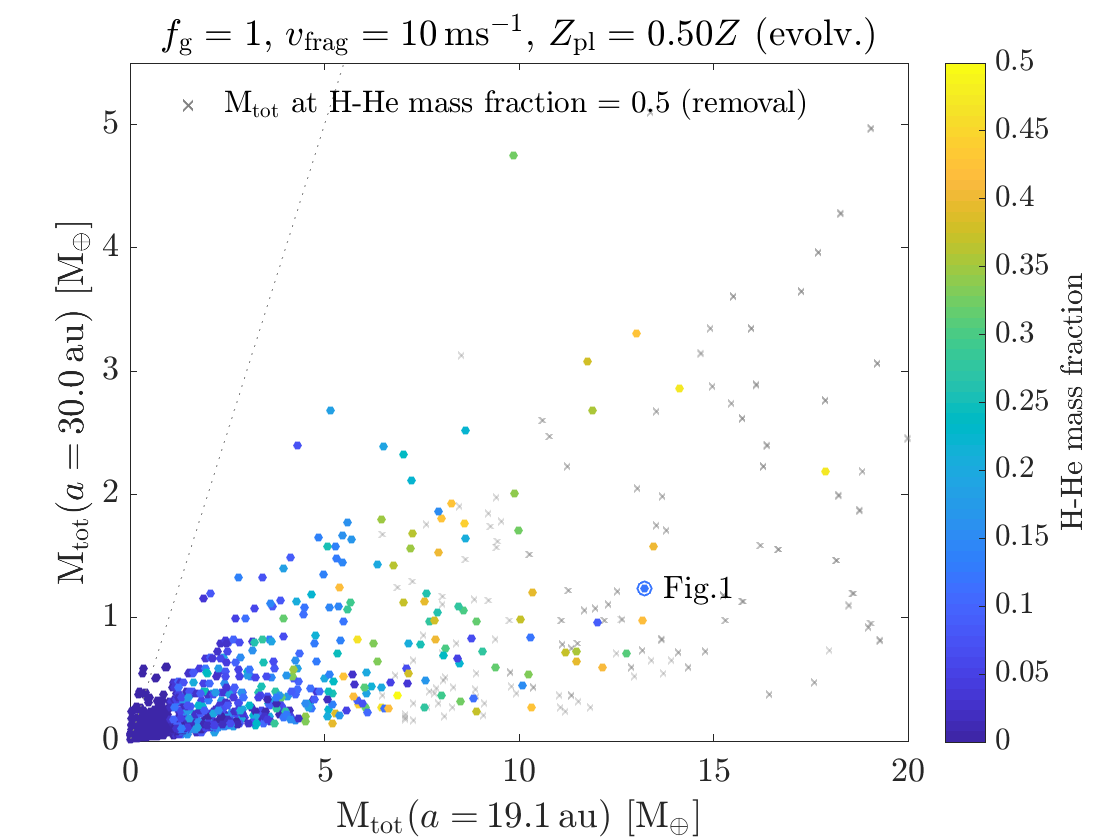}
    \caption{
    Total masses and H-He mass fractions at the time of disk dissipation for all {\it in-situ} simulations with parameters as indicated by the plot title. The grey crosses show simulations where one of the planets obtained a H-He mass fraction above 50\% before disk dissipation. If growth were allowed to continue, most of these planets would grow to become gas giants. The growth-tracks for the simulation marked Fig. 1 is shown in Fig. \ref{fig:1case}. 
    }
    \label{fig:1case_paramstudy}
\end{figure}

In Fig. \ref{fig:1case_paramstudy} we show more detailed results from one of the simulation sets (corresponding to the yellow scatter points in the bottom right panel of Fig. \ref{fig: scatt pebbGas}, in total $\sim 24,000$ simulations). The color of the scatter points represents H-He mass fractions of the planets at $t=t_{\rm disk}$, and the gray crosses indicate the total masses of the planets at the time one of them reaches a H-He mass fraction above 50\%. None of these simulations resulted in a planet of Uranus mass or larger with H-He mass fractions below  20\%. Furthermore, no simulation leads to an outer planet with mass above $5\, \textrm{M}_{\oplus}$, while keeping the  H-He mass fractions of both planets below 50\%. In Fig. \ref{fig:1case_paramstudy}, we find planets with total masses as low as $7\, \textrm{M}_{\oplus}$ that reach H-He mass fractions of 50\%. If allowed, most of these planets would grow to become gas-giants. 

In summary, the results of this section suggest that it is unlikely that Uranus and Neptune formed {\it in-situ} at their current locations, if the embryos formed at the same time and with the same mass. In Section \ref{sect: neptune form earlier}, we investigate whether this result changes when we allow for different embryo masses.

\section{Varying the formation location}\label{sect:vary location}
In the previous section we investigated the possibility of forming both planets at their current locations. However, the initial formation locations of the planets is unknown, and it is likely that some migration took place during and/or after the dissipation of the gas disk. Assuming that both planets formed beyond the current orbit of Saturn, the formation locations could have been somewhere between $\sim 12-40\, \textrm{au}$. The inner boundary comes from requiring dynamical stability with Saturn, and the outer boundary comes from demanding that no planet disturbs the classical Kuiper Belt. We consider the following formation locations for the inner planet: 12, 15, 18, 21, 24, 27 and $30\, \textrm{au}$. We then place the outer planet at 5, 10, 15 and $20\, R_H$ beyond this location, where we used the current mass of Neptune to calculate the Hill radii. This leads to formation locations beyond $40\, \textrm{au}$ in three cases, which we subsequently remove from the study. Due to the complicated nature of planetary migration, it is not  included in our simulations. Instead, we assume that  migration occurred after the dissipation of the gas disk. The possible impacts of planetary migration on our results are discussed in Section \ref{sect: migration}.

We consider successful Uranus and Neptune analogues to have masses within $1.5\, \textrm{M}_{\oplus}$ of their current masses and H-He mass fractions less than 20\%. In Appendix \ref{sect: double mass difference}, we show how the number of successful analogues changes when allowing the planetary masses to differ by $3\, \textrm{M}_{\oplus}$ instead of $1.5\, \textrm{M}_{\oplus}$. 
Since it is possible that Uranus and Neptune have switched places after the dissipation of the disk, we also consider this scenario when identifying successful analogues. In other words, we also consider simulations where the mass of the inner planet is within $1.5\, \textrm{M}_{\oplus}$ of Neptune's current mass, and the mass of the outer planet is within $1.5\, \textrm{M}_{\oplus}$ of Uranus' current mass, to be successful. 

\begin{figure*}
\centering
{\includegraphics[width=\columnwidth]{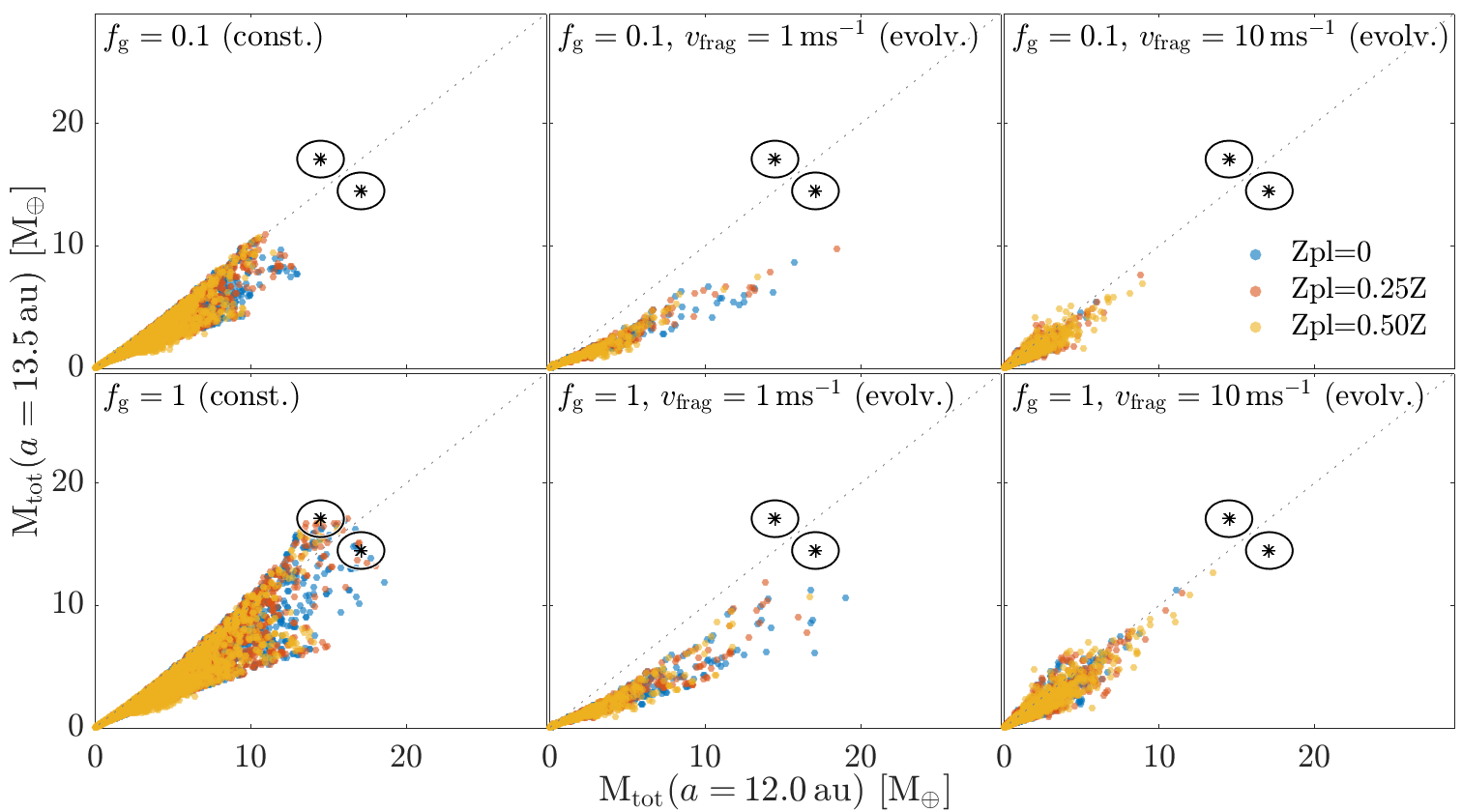}}
{\includegraphics[width=\columnwidth]{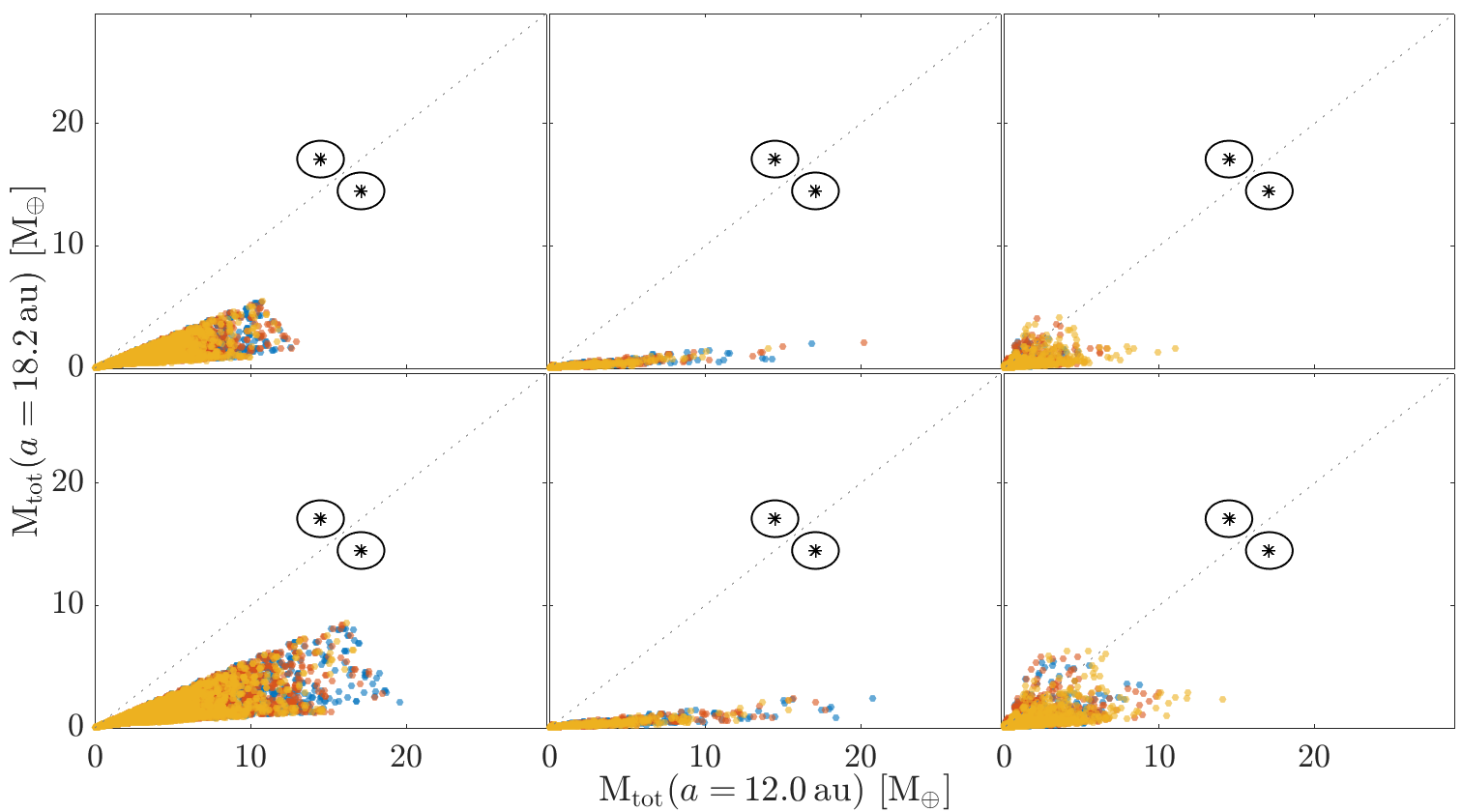}}
{\includegraphics[width=\columnwidth]{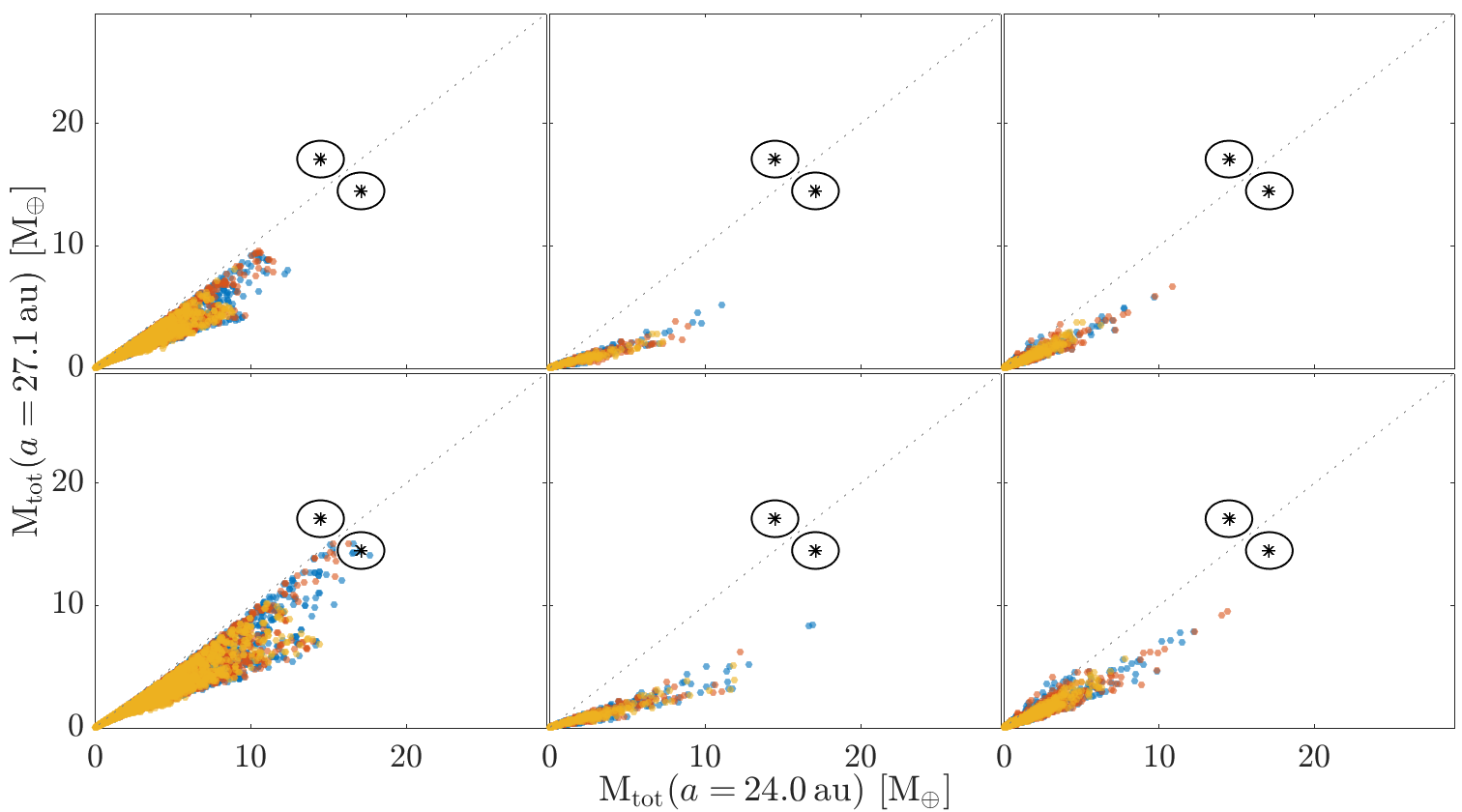}}
{\includegraphics[width=\columnwidth]{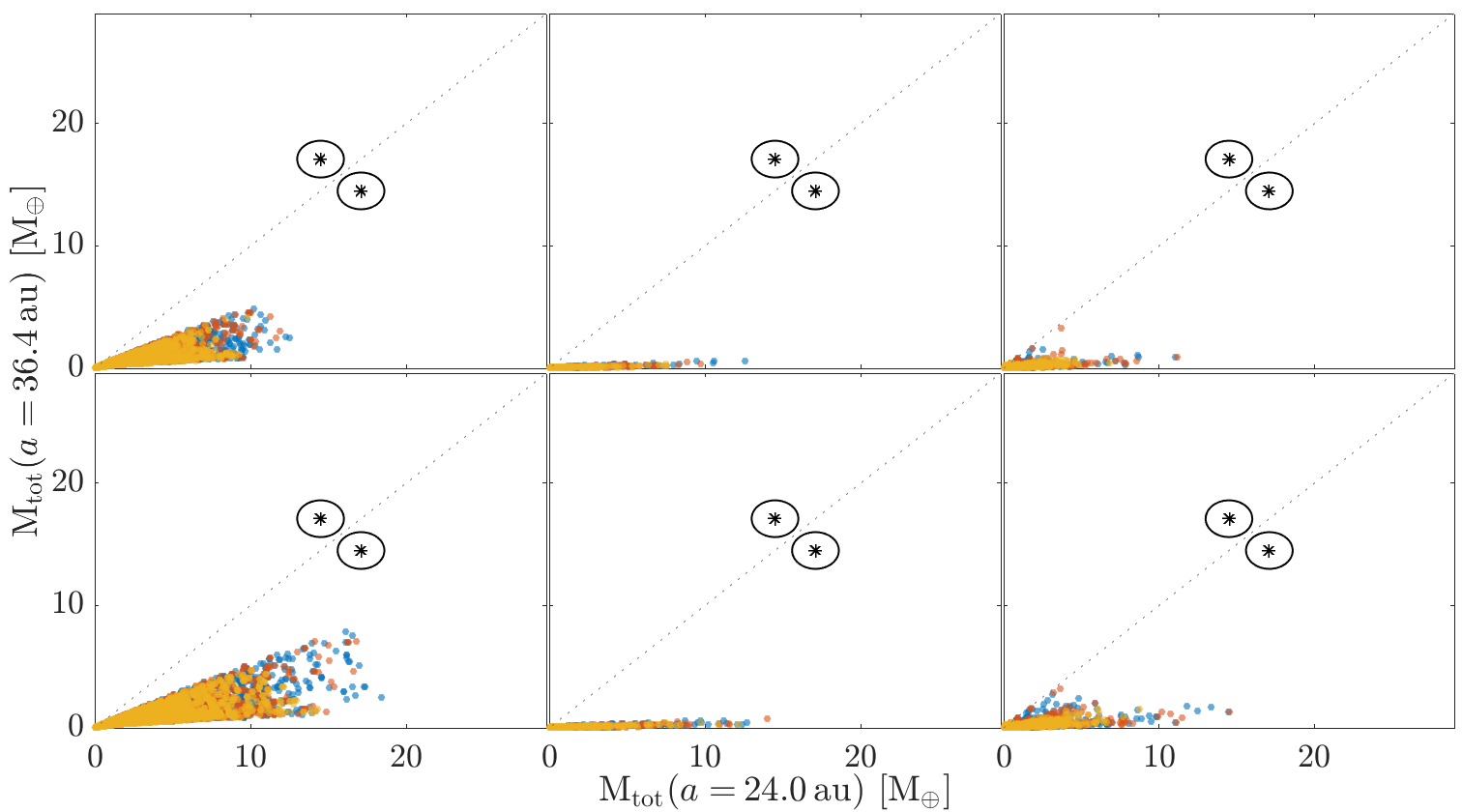}}
\caption{Plots showing the total masses of planets with H-He mass fractions less than 20\%, for simulations with four different semimajor axes configurations (the semimajor axes of the planets are indicated in the x and y-labels). The legends on the top left figure are the same in the other three figures. The top left figure shows results obtained with our most compact configuration, and the bottom right figure shows results obtained with a very non-compact configuration. }
\label{fig: scatt pebbGasPlan vary}
\end{figure*}

Fig. \ref{fig: scatt pebbGasPlan vary} shows a few  examples from the parameter study, where we have used circles to highlight successful analogues. The top left figure shows our most compact configuration, with the inner planet at $12\, \textrm{au}$ and the outer planet at $13.5\, \textrm{au}$. In this scenario, the \texttt{constant} model can produce successful Uranus and Neptune analogues. There are also cases where the blocking of pebbles by the outer planet, leads to the outer planet being more massive than the inner planet. When switching to the more realistic \texttt{evolving} model, we find no successful analogues, since the mass of the outer planet becomes too low in the relevant mass range. In order for these planets to become successful analogues, the outer planet would need to acquire some additional mass via other mechanisms, such as giant impacts (see discussion in Section \ref{sect: giant impacts}). When increasing the orbital distance between the planets, the mass difference increases, resulting in fewer successful analogues. In this scenario, multiple giant impacts would be required to produce Uranus and Neptune analogues. 

\begin{figure*}
\centering
{\includegraphics[width=\columnwidth]{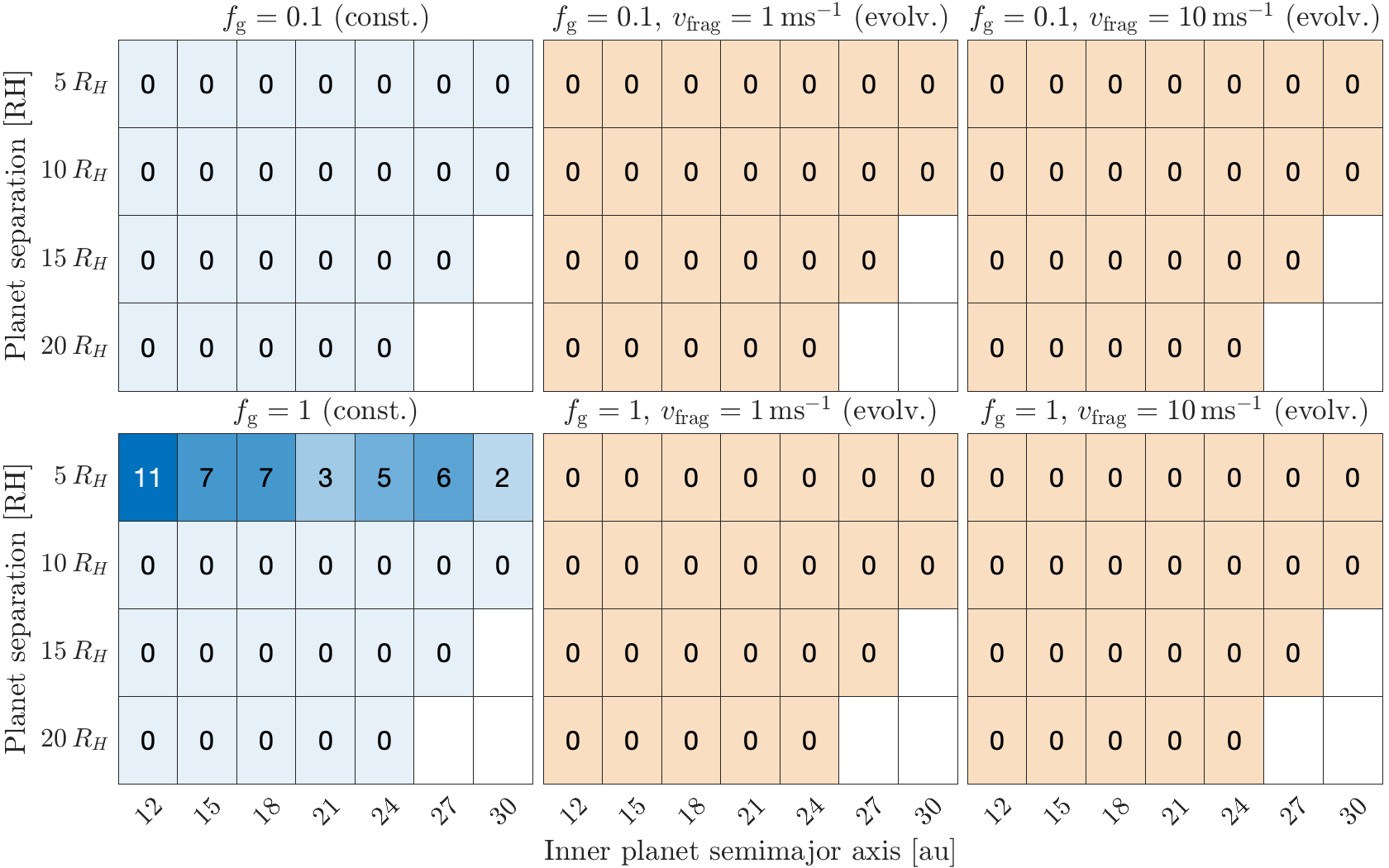}}
{\includegraphics[width=\columnwidth]{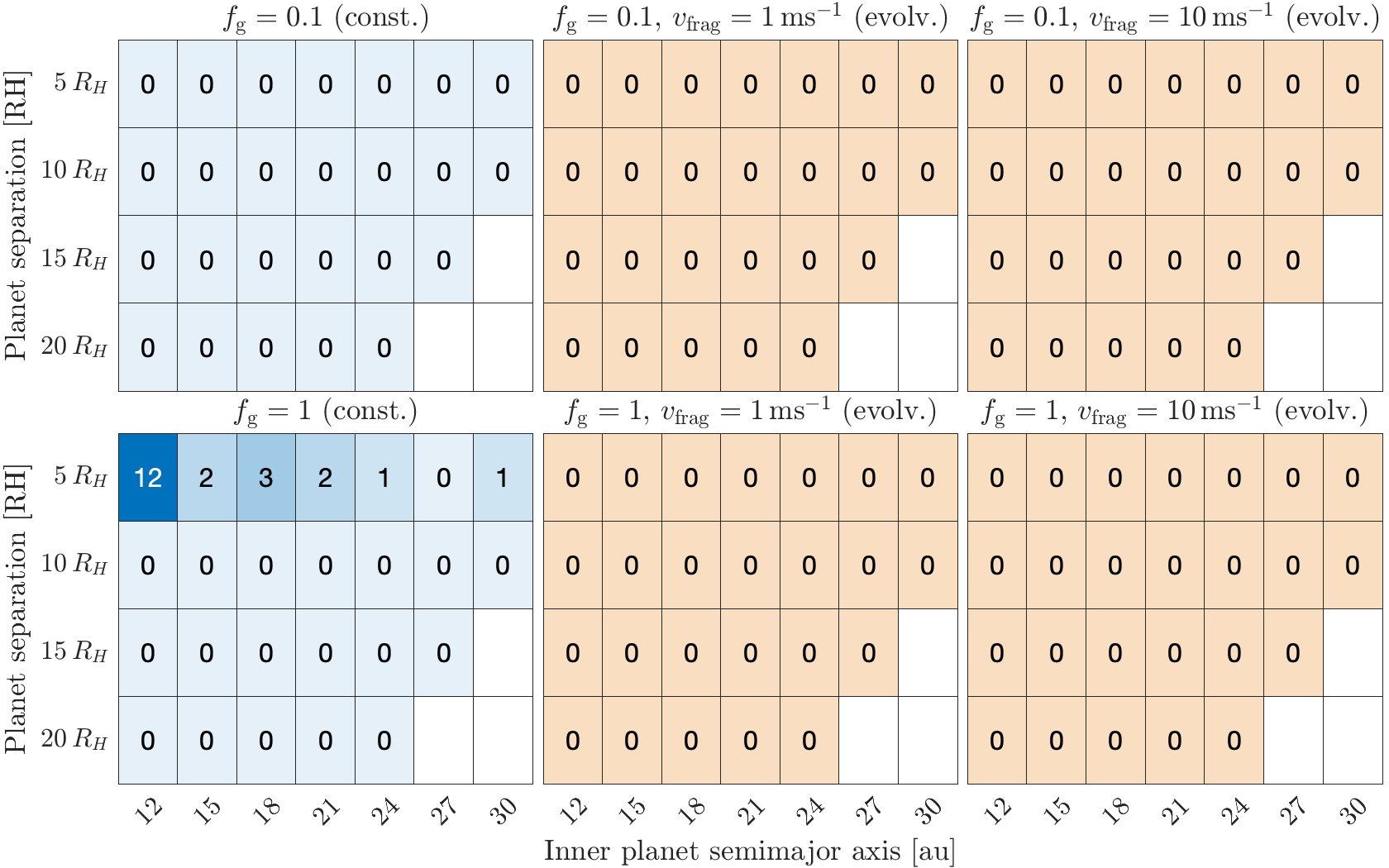}}
{\includegraphics[width=\columnwidth]{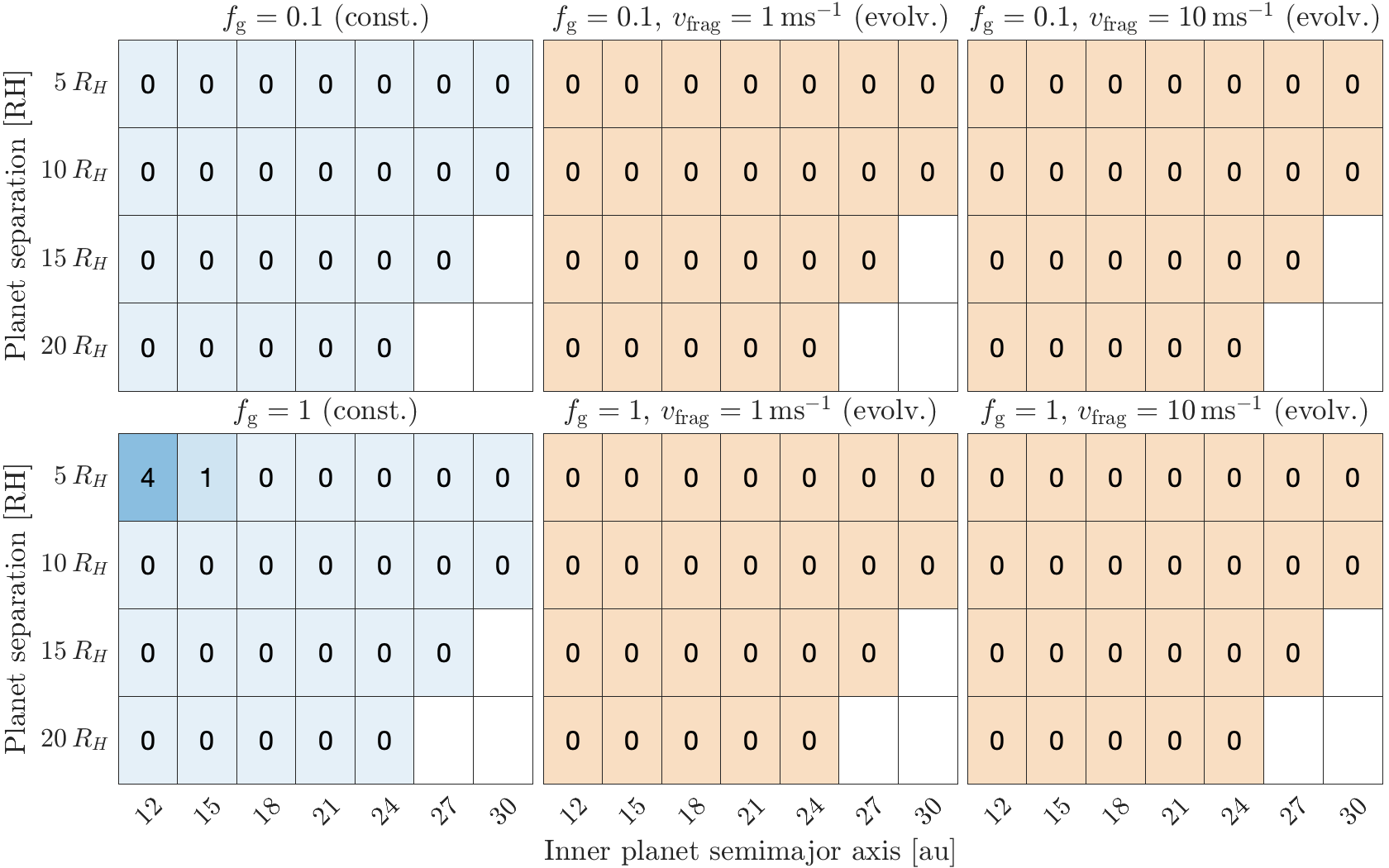}}
\caption{Histograms showing the number of Uranus and Neptune analogues found for each planetary configuration. The top left figure are from simulations with no planetesimal accretion; the top right figure are from simulations with a planetesimal-to-solid mass fraction of 25\%; and the bottom figure are from simulations with a planetesimal-to-solid mass fraction of 50\%. The results show that the \texttt{constant} model (blue histograms) can produce Uranus and Neptune analogues when the separation between the two planets is small. When considering the more realistic \texttt{evolving} model (orange histograms), no analogues are found.}
\label{fig: success pebbGasPlan}
\end{figure*}

The number of successful Uranus and Neptune analogues obtained with each planet configuration is presented in Fig. \ref{fig: success pebbGasPlan}. The number of simulations behind each grid-cell in the histograms is $\sim 80,000$ for the \texttt{constant} model, and $\sim 24,000$ for the \texttt{evolving} model. When using the \texttt{constant} model that has a constant Stokes number and a pebble-flux proportional to the disk accretion rate, we find that successful analogues can be produced when the orbital distance between the planets is equal to 5 Neptune Hill radii. The successful simulations have in common $Z \geq 0.03$, $\dot{M}_{0,\textrm{disk}} \geq 5\times 10^{-8}\, \textrm{M}_{\odot}\textrm{yr}^{-1}$, $t_{\rm start}=1\, \textrm{Myr}$ and $t_{\rm disk}=3\, \textrm{Myr}$. The number of successful analogues decreases with increasing semimajor axes and planetesimal-to-pebble mass fractions.

Our formation model does not lead to  Uranus and Neptune analogues when using the dust-evolution model, regardless of the formation locations that are being used. The main reasons for this are as follows: 1) the H-He mass fractions of planets in the Uranus and Neptune mass range are typically found to be higher than 20\%; and 2) the growth efficiency strongly depends on the semimajor axis, such that a small difference in semimajor axis still generates a relatively large difference in planetary mass. 
Therefore, our results suggest that it is unlikely for Uranus and Neptune to have formed solely via pebble, gas and planetesimal accretion, if the embryos formed at the same time and with the same mass. However, we would like to stress that although we have introduced a rather advanced and comprehensive formation model, simplifications have still been made. For example, we do not consider how the heavy-elements mix within the planetary atmosphere, which could affect the accretion rate of gas onto the planet (see Section \ref{sect: gas acc uncertainties}). We also assume that the initial planetesimal disk is wide-stretched and that the planetesimals are single sized, whereas in reality the surface density of planetesimals could be in-homogeneous within the disk, and the planetesimals are expected to have a size distribution, which is also time dependent. Finally, there are additional processes we have not considered such as giant impacts and planetary migration which could affect our conclusions. We discuss these processes in the following section.  

\section{Potential solutions}\label{sect:potential solutions}
We consider the growth of Uranus and Neptune via pebble, gas and planetesimal accretion. We can form both planets in our simulations when we assume a constant Stokes number, a pebble flux that is proportional to the disk accretion rate and a small orbital separation between the planets. 
The first two assumptions are often used in pebble accretion simulations; but as we have demonstrated, the growth via pebble accretion can significantly change  when considering a more realistic model where the dust population evolves with time and semimajor axis. When we switch to using the \texttt{evolving} model, we no longer find any successful Uranus and Neptune analogues, despite the large range of formation locations and parameters that were used. In this section we discuss potential mechanisms that could assist in forming Uranus and Neptune simultaneously. 

\subsection{What if the outer planet had a head start?}\label{sect: neptune form earlier}
In our main study, we assumed that the embryos of Uranus and Neptune formed at the same time and with the same mass. However, since the efficiency of pebble accretion strongly depends on the semimajor axis and the planetary mass, this leads to the outer planet being much less massive than the inner one. This mass difference could significantly decrease if the outer embryo formed earlier than the inner one, or similarly, if the outer embryo is more massive than the inner one. Since the process of planet formation is rather stochastic, it is certainly possible that embryos form at different times and with different masses.  

\begin{figure*}
\centering
{\includegraphics[width=1.4\columnwidth]{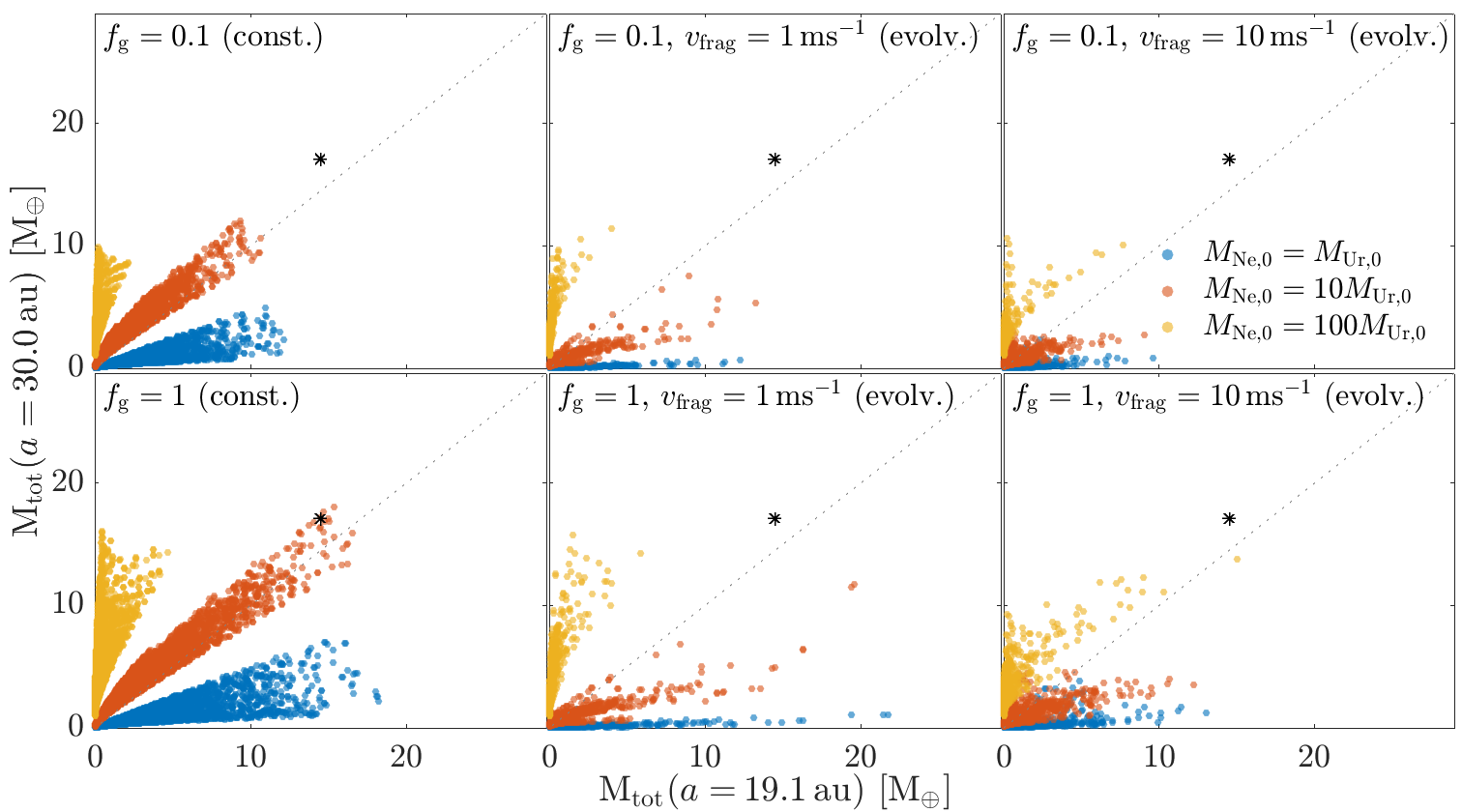}}
\caption{Total mass of planets that are forming {\it in-situ} and has a H-He mass fraction less than 20\% at the time of disk dissipation. Blue scatter points show the results when the planetary embryos have the same mass; red scatter points show the results when the outer embryo is 10 times more massive than the inner one; and yellow scatter points show the results when the outer embryo is 100 times more massive than the inner one. These simulations are performed without planetesimal accretion. }
\label{fig: scatt pebbGas nLTu}
\end{figure*}

\begin{figure*}
\centering
{\includegraphics[width=\columnwidth]{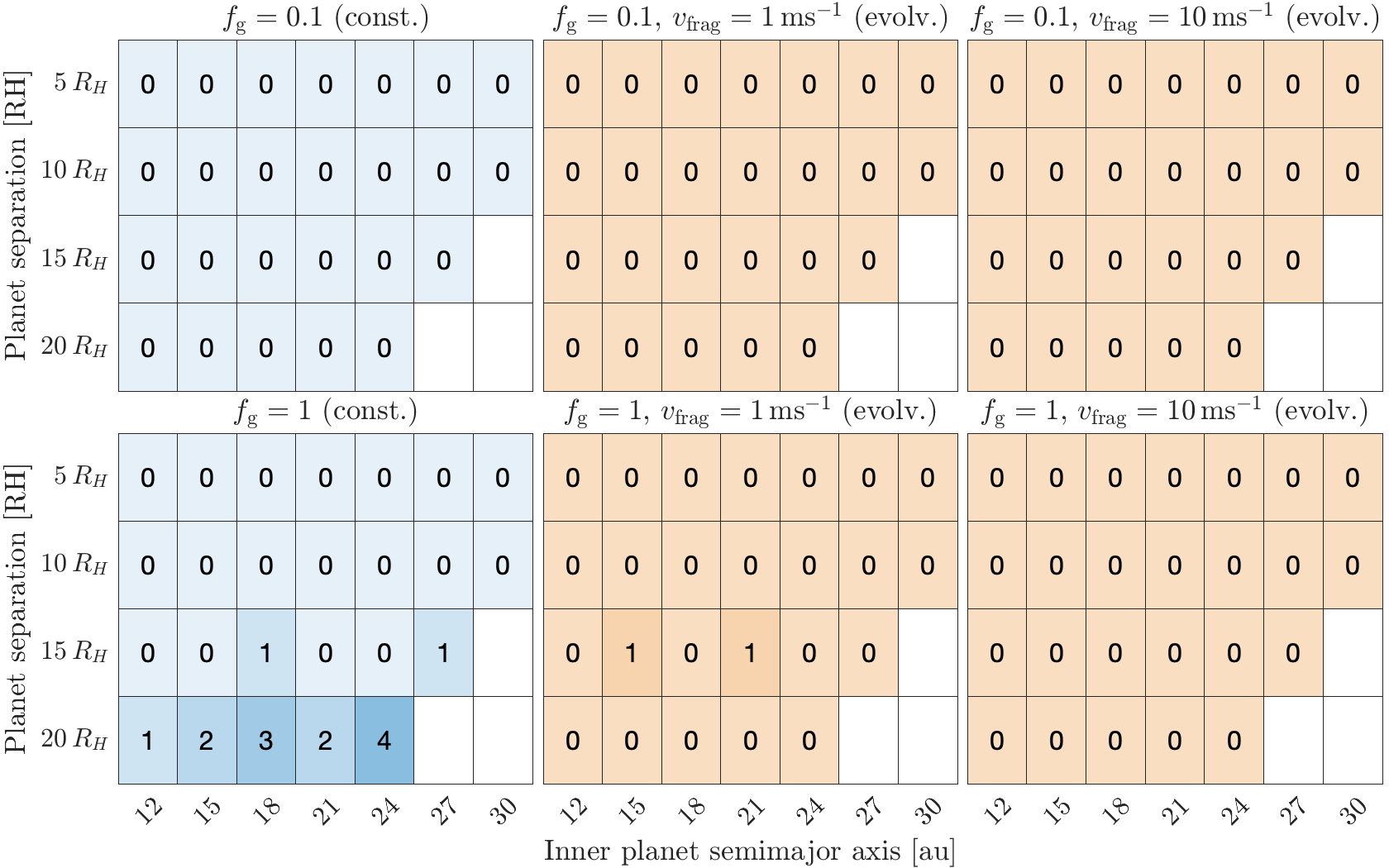}}
{\includegraphics[width=\columnwidth]{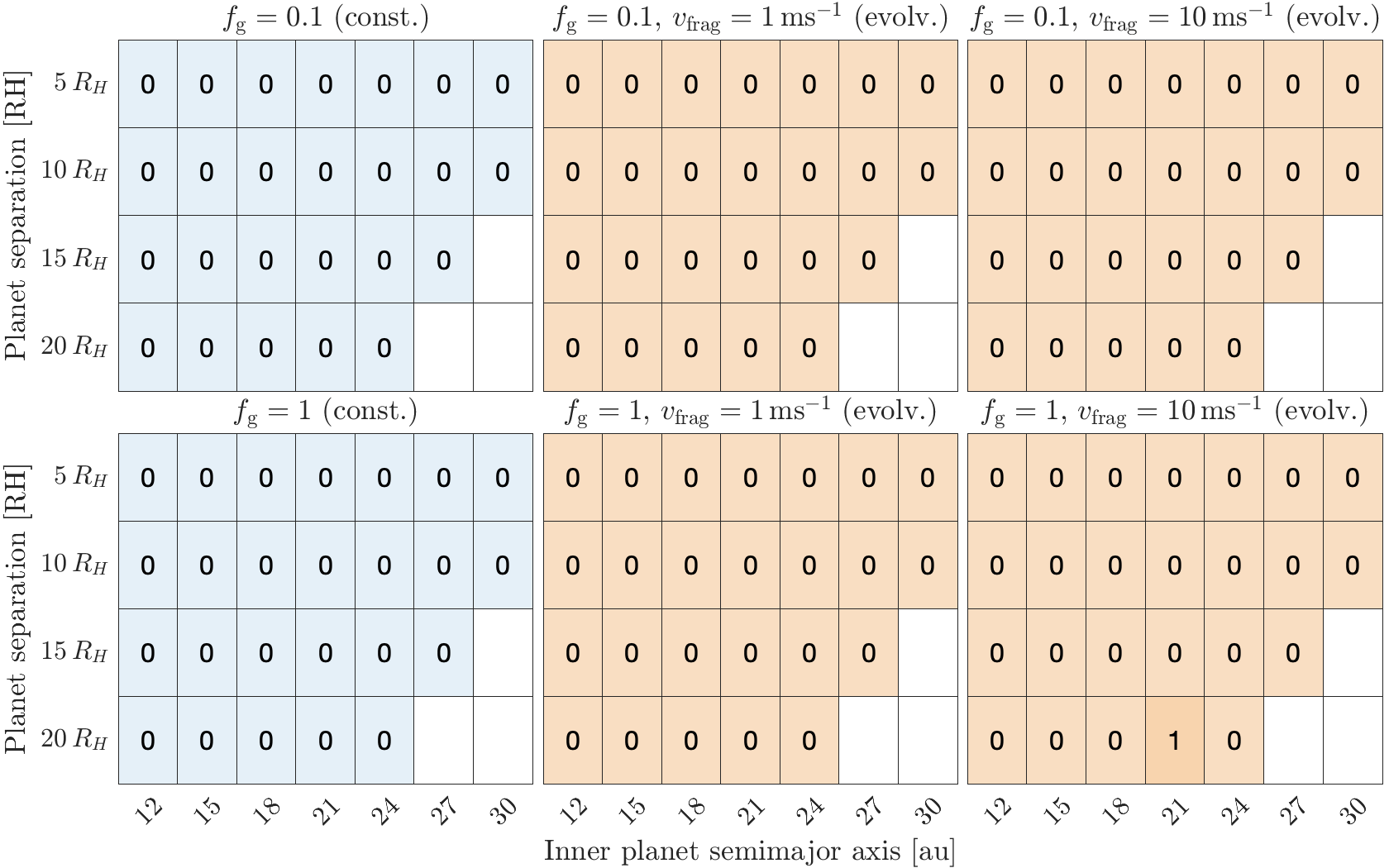}}
\caption{Histograms showing the number of Uranus and Neptune analogues that are found when the outer embryo is 10 times more massive than the inner embryo (left figure), and when the outer embryo is 100 times more massive than the inner one (right figure). The simulations are performed without planetesimal accretion. In contrast to when the embryos had the same mass (Fig. \ref{fig: success pebbGasPlan}), we now produce a few analogues while using the \texttt{evolving} model.}
\label{fig: success pebbGas nLTu}
\end{figure*}

We perform additional simulations where the mass of the outer embryo is increased by a factor of 10 and 100 compared to the inner embryo (with a mass of $0.01\, \textrm{M}_{\oplus}$). This difference in mass could be due to an earlier embryo formation and/or a more massive embryo being formed. We limit this study to the case when all solid accretion occurs via pebble accretion ($Z_{\rm pl}=0$). Fig. \ref{fig: scatt pebbGas nLTu}, shows how the results for the case of {\it in-situ} formation change when the mass of the outer embryo is increased. The number of successful analogues that are obtained at various formation locations are shown in Fig. \ref{fig: success pebbGas nLTu}.  

When considering {\it in-situ} formation, the \texttt{constant} model with the default grain opacity produces Uranus and Neptune analogues when the outer embryo is 10 times more massive than the inner one. When the difference in embryo mass is further increased, the outer planet becomes too massive compared to the inner one. When we use the more realistic \texttt{evolving} model, we do not obtain any successful analogues; however, we are much closer to doing so than we were when we used embryos of the same mass.

When we vary the formation locations, we find a few successful Uranus and Neptune analogues, both while using the \texttt{constant} model and the more realistic \texttt{evolving} model. The successful analogues have in common $\textrm{R}_{\rm edge} \geq 100\, \textrm{au}$, $Z \geq 0.03$, $\alpha_{\rm T} < 10^{-4}$, $\dot{M}_{0,\textrm{disk}} \geq 5\times 10^{-8}\, \textrm{M}_{\odot}\textrm{yr}^{-1}$ and $t_{\rm disk}=3 \times 10^6\, \textrm{yr}$. Unlike when the embryos formed with the same mass, the required separation between the planets is now typically around $15-20\, \textrm{R}_{\rm H}$, rather than $5\, \textrm{R}_{\rm H}$. If we are more generous when searching for successful analogues, and allow the mass difference to be $3\, \textrm{M}_{\oplus}$ instead of $1.5\, \textrm{M}_{\oplus}$, the number of successful analogues is more than doubled (see Appendix \ref{sect: double mass difference}). The parameters leading to successful analogues varies more in this scenario, but they still have in common $t_{\rm disk}=3 \times 10^6\, \textrm{yr}$. This is the case for all successful analogues found throughout the study, and the reason is that when the disk lifetime is longer, continuous gas accretion results in H-He mass fractions above 20\%.

{\it In summary, we can simulate the formation of Uranus and Neptune solely via pebble, gas and planetesimal accretion if the outer planetary embryo forms earlier and/or more massive than the inner one}. The required disk metallicity is above 1\%, and overall, very specific parameters are required in order to form Uranus and Neptune. Therefore, the formation of Uranus and Neptune remains a challenge to planet formation theories. 

\subsection{What if the planets migrated?}\label{sect: migration}
The effects of disk-driven migration on the formation of Uranus and Neptune are hard to predict, since the migration of the embedded planets can be directed both inwards and outwards depending on the surrounding disk conditions \citep{Paardekooper2022}. Let us assume that the planets migrate inwards during formation via classical Type-I migration (see e.g. eq. 3-4 of \citealt{Johansen2019}). If the negative radial gradients of gas surface density and temperature are $15/14$ and $3/7$ respectively, and we assume an unperturbed gas surface density, the migration rate has a weak negative dependence on semimajor axis and is directly proportional to the planetary mass. Since in our models the inner planet typically grows faster than the outer planet, the inner planet would thus migrate faster inwards than the outer planet. This would lead to a faster growth for the inner planet, since pebble accretion is more efficient at smaller semimajor axes, leading to an even larger inward migration, etc... 
In that case, the effect of migration would enhance the mass difference between the planets in our simulations, making it even more challenging to form Uranus and Neptune. 

In the above line of argument we assumed an unperturbed disk. If the planets open up a gap in the gas surface density profile, it could affect the migration rate (see e.g. \citealt{Kanagawa2018}). However, since a planet beyond $10\, \textrm{au}$ with a mass $<20\, \textrm{M}_{\oplus}$ is not expected to open up a deep gap, our assumption should be justified. Torques exerted by the dust disk and significantly different temperature and surface density structures could change the migration rate compared to the classical Type-I rate. The migration of planets can further be halted by trapping in resonances with other planets.

Planetary migration is further expected to affect the accretion efficiency of planetesimals. However, it is difficult to assess the magnitude of this effect. 
A migrating planet can increase the surface density of planetesimals $\Sigma_{\rm pl}$ in its vicinity, because planetesimals that are initially outside the feeding zone can enter the feeding zone \citep{Tanaka_1999,Alibert_2005,Shibata_2020,Turrini_2021,Shibata_2022a}.
However, mean motion resonances of a massive protoplanet like Jupiter were found to prevent planetesimals from entering the planet's feeding zone \citep{Shibata_2020,Shibata_2022a}, which reduces $\Sigma_{\rm pl}$.
Mean motion resonances of a Neptune-size planet are not as strong as those of a massive giant planet, but it could still change the planetesimal accretion rate. 
In addition, gravitational scattering of other protoplanets can affect the configuration of mean motion resonances \cite[e.g.][]{Tanaka_1997,Levison_2010}.
The mutual gravitational interaction of the forming Uranus and Neptune could affect the planetesimal accretion rate if they experienced radial migration. 

 To summarise, it is hard to predict how disk-driven migration would have affected the formation of Uranus and Neptune, but it is reasonable to assume for some sort of disk-driven migration to have taken place.   

\subsection{Giant impacts}\label{sect: giant impacts}
Our results show that it is very hard to form Uranus and Neptune simultaneously, and in the few cases where we do succeed, the required disk metallicity is above the typically assumed 1\%. 
The main challenge concerns keeping the H-He mass fractions and mass difference between the two planets small. One possible solution is giant impacts. The H-He mass fractions of lower mass planets are typically below 20\%, and Uranus and/or Neptune could have formed by colliding two or more such planets together. For example, in the cases where we form an Uranus analogue, but the corresponding mass of Neptune is too low, a giant impact onto Neptune could deliver the missing heavy-element  mass while not increasing the H-He mass fraction. The likelihood for such an event to have taken place that far out in the Solar System is unknown, although collisions are expected to be common \citep[e.g.,][]{2015A&A...582A..99I,2021MNRAS.502.1647C}. 

\subsection{The effect of disk substructures}
In our model we assume that the gaseous disk is smooth and evolves as a viscous accretion disk. However, observations of protoplanetary disks have revealed that many disks harbour substructures \citep[e.g.,][]{Andrews2018}. The most commonly observed substructure in the dust distribution is rings and gaps, where several observed rings have been shown to be consistent with dust trapping inside pressure maxima \citep[e.g.,][]{Dullemond2018}. Such pressure maxima could be the result of some fluid dynamics process, icelines or planet-disk interactions (see \citealt{Bae2023} for a recent review on disk substructures).

Since the surface density and radial drift of pebbles in a structured disk differ considerably compared to a smooth disk \citep{Eriksson2020}, the growth of planets via pebble accretion would also be affected.  Similarly, local variations in the temperature and gas surface density structure could affect the direction and speed of migration, and result in migration traps \citep{GuileraSandor2017}. Because pressure maxima collects large amounts of pebbles, they can be efficient sites for planetesimal formation and further planetary growth (\citealt{Guilera2020,Chambers2021,Lau2022,JiangOrmel2023}).

Indeed substructures in the disk would affect the formation history of the growing planets. 
If Uranus and/or Neptune were to form at the location of a pressure maxima, their growth could occur at much shorter time-scales than inferred in this work.  
At the same time, the pebble flux  interior to a pressure maxima could also be significantly smaller,  potentially leading to slower growth for planets that reside closer to the star than the pressure maxima. The effect of disk substructure on the growth of Uranus and Neptune is  complex and should be investigated in detail in future research. 

\subsection{Gas accretion rate uncertainties}\label{sect: gas acc uncertainties}
Our calculations of the gas accretion rates are self-consistent and represent a significant improvement in comparison to the commonly used simplifying assumption that gas accretes as a constant fraction of the solid accretion rate. Nevertheless,  there are several simplifications in our gas accretion model that could be improved in future work. For example, our models do not include the interaction of the solid material in the envelope. Pebbles and planetesimals, whether they are made of rock or ice are expected to vaporize and enrich the envelope with heavier elements  \citep[e.g.,][]{Pollack_1986, Podolak_1988}. This pollution of the envelope with heavy elements can accelerates the planetary growth, making giant planet formation more efficient \citep[e.g.]{Stevenson_1982, Hori_2011, Venturini_2015, Valletta_2020}. If this is the case, forming Uranus and Neptune would be more difficult in our current model.

However, this mechanisms is not completely understood and has received attention primarily in the context of Jupiter's  formation. 
It remains possible that a more realistic gas accretion prescription actually leads to lower gas accretion rates. This is suggested from three-dimensional gas accretion models that are found to have  lower gas accretion rates than one-dimensional models \citep{Ormel_2015, Cimerman_2017}, although this is predicted for planets forming at much shorter orbital periods than Uranus and Neptune (e.g. \citet{Moldenhauer_2021, Moldenhauer_2022} calculated recycling rates at 0.1 AU). Pebble enrichment of the nebular gas in the locations were the planets form  could also contribute to recycling \citep{Wang_2023}. 
It is therefore clear that inferring realistic gas accretion rates for planets forming at large orbital distances, such as Uranus and Neptune, is desirable. Currently, it is unknown how such enrichment could affect the formation of Uranus and Neptune.

\subsection{Planetesimal accretion rate uncertainties}\label{sect: planetesimal acc uncertainties}
To calculate the planetesimal accretion rate, we adopt the statistical model. 
However, this model has been developed for modelling the formation of terrestrial planets and cores of giant planets.
The effect of the gas accretion is not included in the current statistical model.
For example, \citet{Shibata_2023} showed that the current statistical approach cannot reproduce the results of N-body simulations once the protoplanet enter the runaway gas accretion.
Uranus and Neptune have not entered the runaway gas accretion, 
but the steady gas accretion during the planetary growth could affect the planetesimal accretion rate, and we hope to address this in future research. 

\section{Conclusions}\label{sect:conclusions}
We have studied the formation of Uranus and Neptune via pebble, gas, and planetesimal accretion. 
We considered two different models for pebble accretion: a simple model with constant Stokes number and a pebble flux proportional to the disk accretion rate; and a more realistic model where the Stokes number and pebble flux are obtained from a dust-evolution model and vary with time and semimajor axis. We do not include migration, but test a wide range of formation location and a wide range of disk parameters. Our main conclusions can be summarised as follows:

\begin{itemize}
    \item If the embryos form at the same time and with the same mass, our formation model with an evolving dust population is unable to produce Uranus and Neptune analogues, regardless of the assumed formation locations. When we use the simpler and less realistic \texttt{constant} model, Uranus and Neptune analogues can form when the orbital distance between the planets is small. 
    \item If the outer embryo forms earlier and/or more massive than the inner embryo, we can form both planets simultaneously in a few instances where the disk is metal-rich and has a lifetime of a few Myr. Overall, very specific parameters are required to form Uranus and Neptune, and the formation of these planets remains a challenge to planet formmation theory. 
    \item Based on our results, it is unlikely for Uranus and Neptune to have formed \textit{in-situ}. When we use the \texttt{evolving} model, we do not produce any analogues regardless of the assumed formation locations and embryo masses. In the \texttt{constant} model, we find Uranus and Neptune analogues when the outer embryo is 10 times more massive than the inner embryo.  
    \item The key challenge in forming Uranus and Neptune is keeping the H-He mass fractions below 20\% and keeping the planetary masses similar.  When the grain opacity is low or the disk lifetime is longer than $\sim 3\, \textrm{Myr}$, the H-He mass fractions become too large in the mass regime relevant for Uranus and Neptune. The mass difference between the planets increases with the orbital separation between the planets, since the pebble accretion rate strongly depends on the semimajor axis. 
\end{itemize}


Our study demonstrates the complexity of modelling planet formation properly. In addition, it clearly shows how different model assumptions affect the results concerning the forming planets. 
This is not only important for improving our understanding of the origin of Uranus and Neptune, but also of the origin of the many intermediate-mass exoplanets detected in our galaxy.  

Finally, we suggest that more accurate determinations of the H-He mass fractions in Uranus and Neptune would be valuable in constraining their formation path. We therefore look forward to a future mission to Uranus and Neptune as well as to the upcoming accurate measurements of mass, radius, and atmospheric compositions of intermediate-mass/size exoplanets. 

\section*{Acknowledgements}
The authors wish to thank the anonymous referee for providing helpful comments that lead to an improved manuscript. LE acknowledges funding by the Institute for Advanced Computational Science Postdoctoral Fellowship. LE would further like to thank Stony Brook Research Computing and Cyberinfrastructure, and the Institute for Advanced Computational Science at Stony Brook University for access to the SeaWulf computing system, made possible by grants from the National Science Foundation (\#1531492 and Major Research Instrumentation award \#2215987), with matching funds from Empire State Development’s Division of Science, Technology and Innovation (NYSTAR) program (contract C210148).\\
MML, SS and RH acknowledge that part of this work has been carried out within the framework of the National Centre of Competence in Research PlanetS supported by the Swiss National Science Foundation under grants 51NF40\_182901 and 51NF40\_205606. The authors acknowledge the financial support of the SNSF.
  RH and SS also  acknowledge support from SNSF under grant \texttt{\detokenize{200020_188460}}. 
\section*{Data Availability}
The data underlying this article will be shared on reasonable request to the corresponding author.



\bibliographystyle{mnras}
\bibliography{refs} 



\appendix

\section{Dust evolution with the \texttt{pebble predictor}}\label{sect:pebble predictor}
The \texttt{pebble predictor} code was presented in \citet{Drazkowska2021}, and is publicly available at https://github.com/astrojoanna/pebble-predictor. It is a semi-analytic model for predicting the flux-averaged Stokes number and total flux of pebbles at all locations and at all times in an arbitrary unperturbed disk. The \texttt{pebble predictor} takes as input the initial radial surface density profiles of gas ($\Sigma_{\rm disk,0}$) and dust ($\Sigma_{\rm dust,0}=Z \Sigma_{\rm disk,0}$), the radial temperature structure ($T$), turbulent strength ($\alpha_{\rm T}$), internal density of dust grains ($\rho_{\bullet}$) and fragmentation velocity ($v_{\rm frag}$). The initial size of the dust grains is assumed to be $1\, \mu \textrm{m}$, and growth proceeds via turbulence driven collisions until either the fragmentation or radial drift barrier is reached. The time evolution of the gas disk is neglected in the \texttt{pebble predictor} model, and the evolution of the dust disk is approximated by keeping track of the dust movement. Despite these simplifications, the results of the \texttt{pebble predictor} can reproduce rather well the outcome of full coagulation simulations (see Fig. 7 in \citealt{Drazkowska2021}). 

\begin{figure*}
\centering
{\includegraphics[width=\columnwidth]{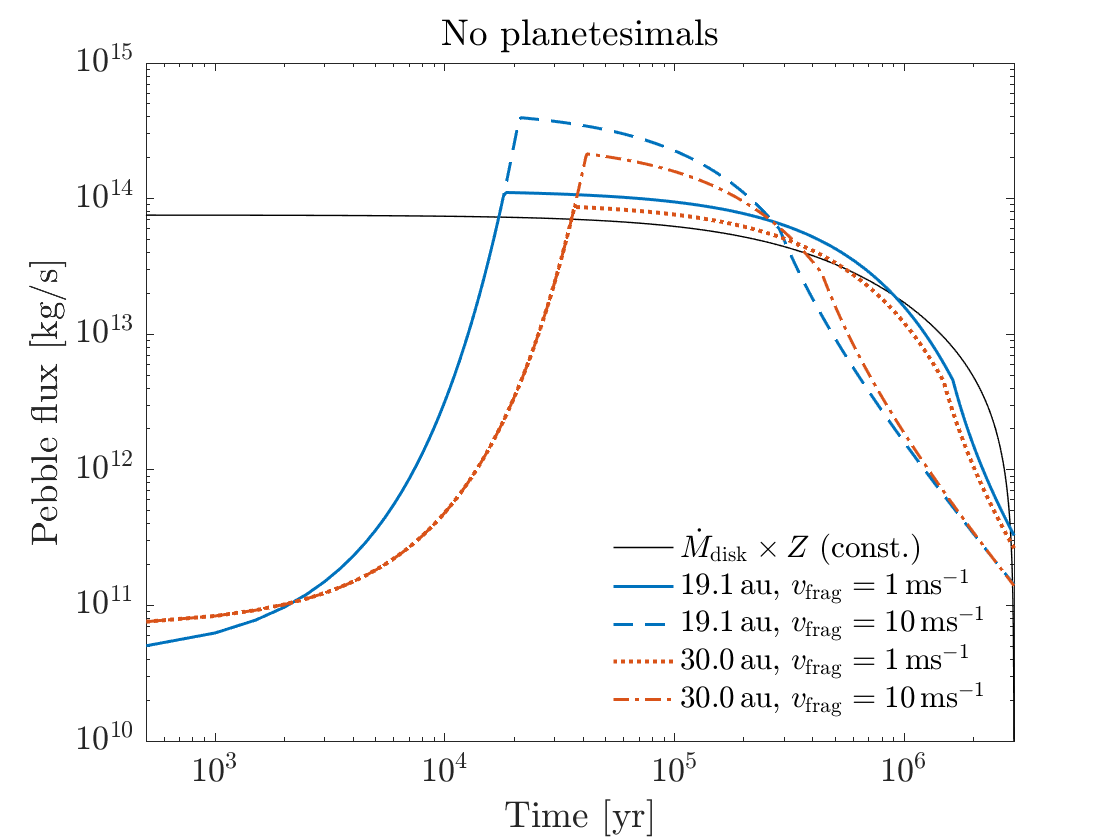}}
{\includegraphics[width=\columnwidth]{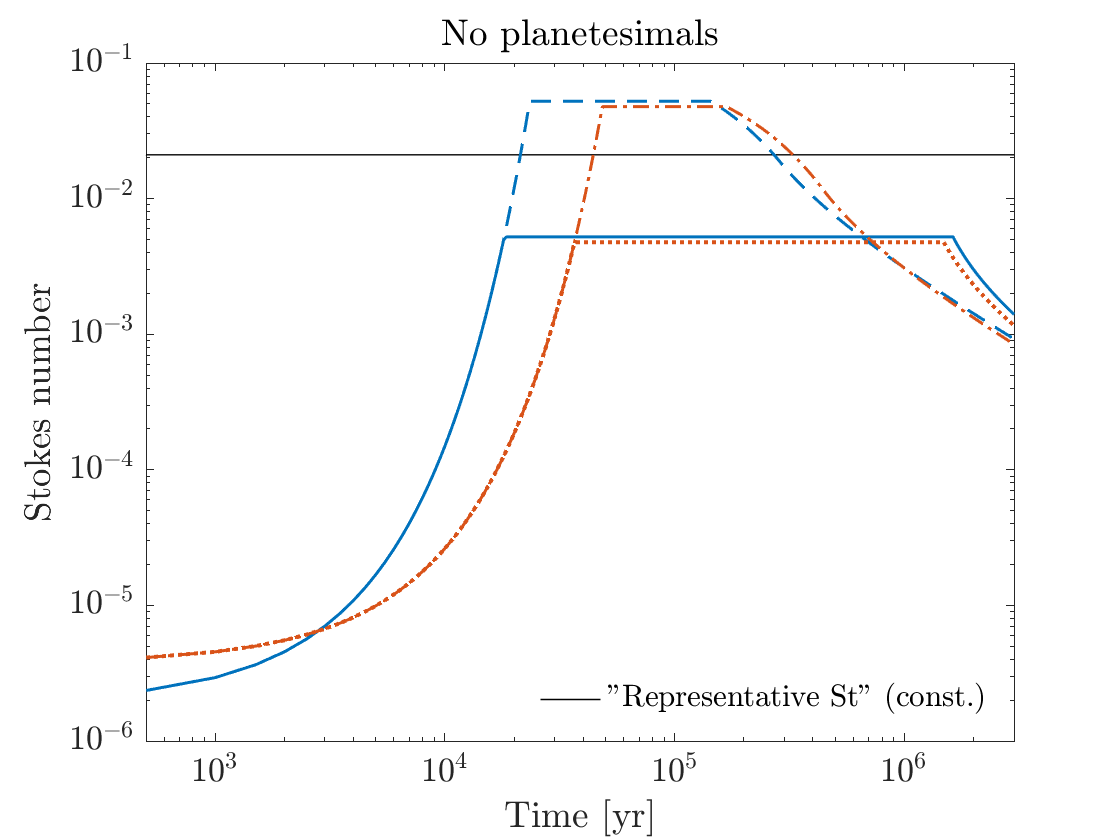}}
{\includegraphics[width=\columnwidth]{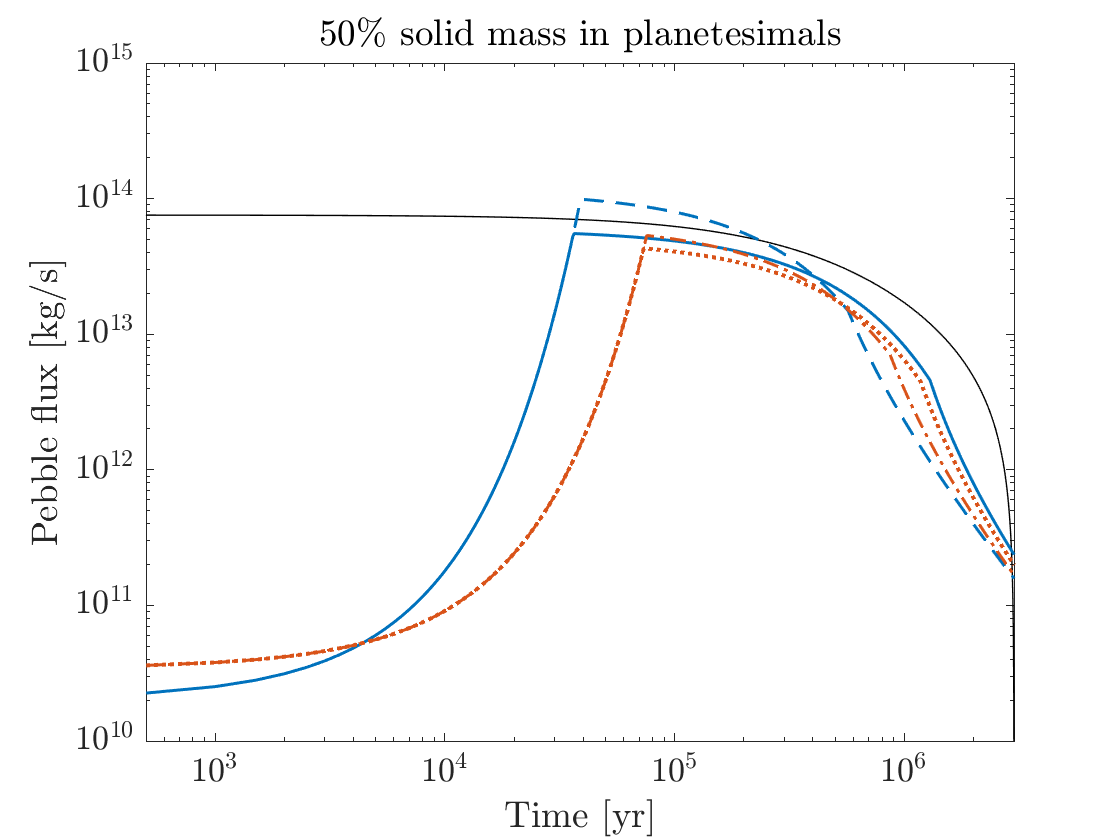}}
{\includegraphics[width=\columnwidth]{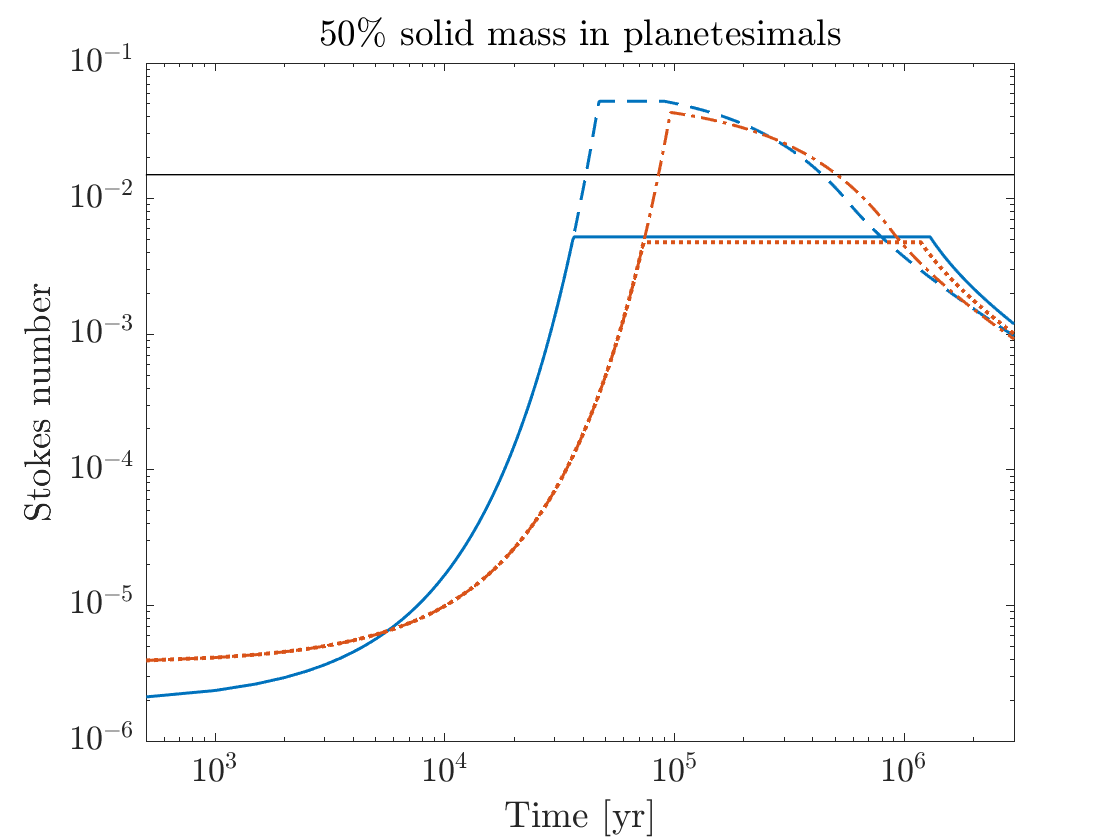}}
\caption{The time evolution of the pebble flux (left) and Stokes number (right) that is obtained from the \texttt{pebble predictor} when using the parameters given in Table \ref{tab: 2 cases parameters}. The pebble flux obtained from taking the product of the disk metallicity and the disk accretion rate (the \texttt{constant} model) is shown with black lines in the left panels. The black lines in the right panels show a "representative" Stokes number (see text for details).}
\label{fig: 2 cases flux stokes}
\end{figure*}

We consider two different fragmentation velocities, $v_{\textrm{frag}}=1\, \textrm{m}s^{-1}$ and $10\, \textrm{m}s^{-1}$, and assume $\rho_{\bullet}=1000\, \textrm{kg}\textrm{m}^{-3}$. We use a logarithmic radial grid with inner edge at $1\, \textrm{au}$, outer edge at $50$, $100$ or $200\, \textrm{au}$, and 100, 200 or 400 grid points, respectively. The time grid has 10,000 logarithmically spaced grid points and stretches from $1\, \textrm{yr}$ to $10\, \textrm{Myr}$. In Fig. \ref{fig: 2 cases flux stokes}, we show the time evolution of the pebble flux and Stokes number at the current semi-major axes of Uranus and Neptune, for the set of disk parameters given in Table \ref{tab: 2 cases parameters}. We further show how the results vary with the assumed fragmentation velocity, and how they change when considering that half of the solid mass is locked up in planetesimals (bottom panels). 

When we use the higher of the two fragmentation velocities, the pebbles grow to significantly larger sizes. The corresponding pebble flux also peaks at higher values; however, because of this the pebble reservoir is also depleted more quickly. When comparing to the pebble flux that results in from taking the product of the disk metallicity and the disk accretion rate (the \texttt{constant} model, black lines), the \texttt{pebble predictor} in this case results in lower pebble fluxes at late times. In the \texttt{constant} model, we use a Stokes number that is constant with time and semi-major axes. The black line in the right panels show a "representative Stokes number", which is obtained after: calculating the time-average of the flux-weighted Stokes number; and taking the average of the resulting values obtained at the two semi-major axes and with the two fragmentation velocities. We use this representative Stokes number to compare the growth tracks obtained with the \texttt{constant} model and the \texttt{evolving} model in Appendix \ref{sect: 2 selected cases}. 

\section{Growth-tracks from an example simulation}\label{sect: 2 selected cases}
\begin{figure*}
\centering
{\includegraphics[width=\columnwidth]{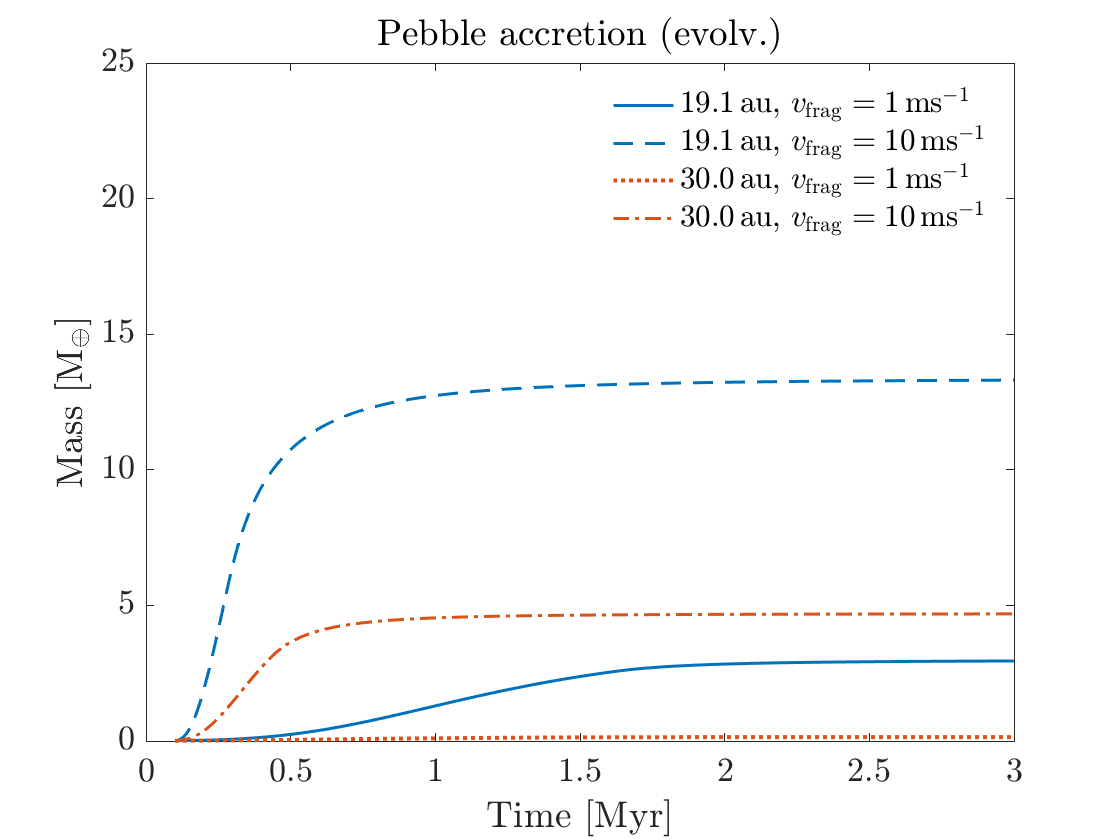}}
{\includegraphics[width=\columnwidth]{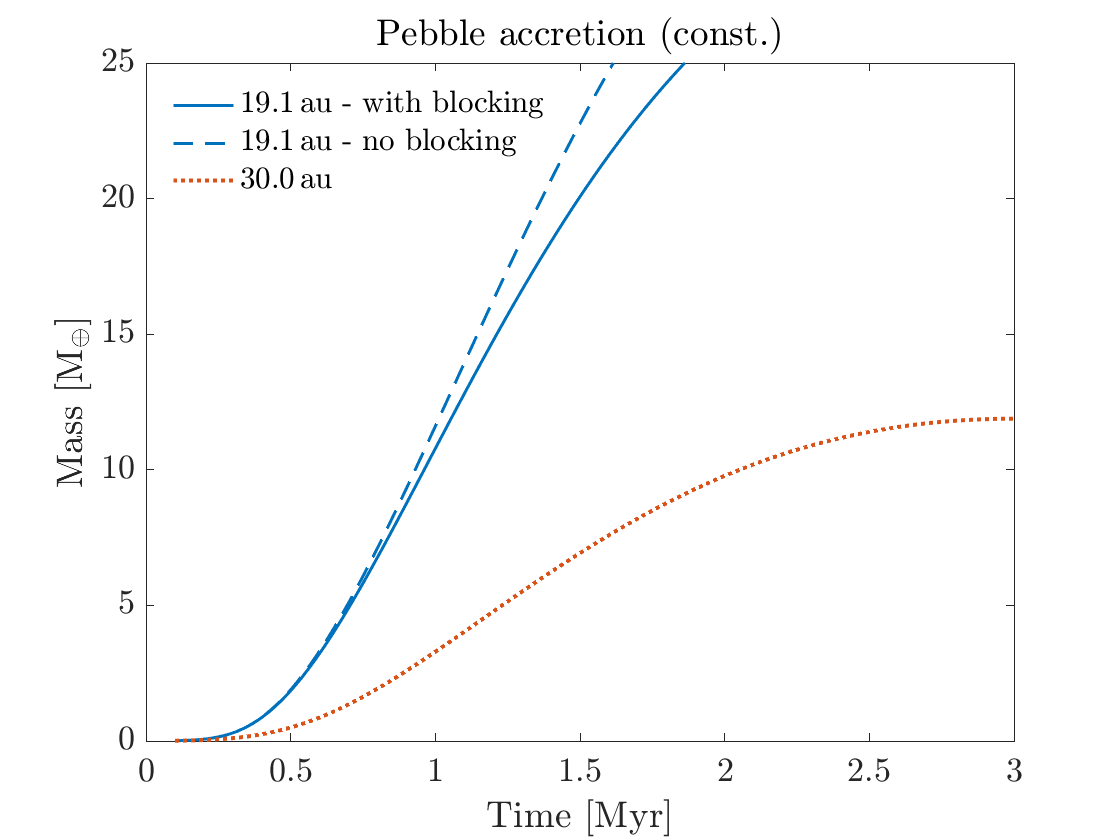}}
{\includegraphics[width=\columnwidth]{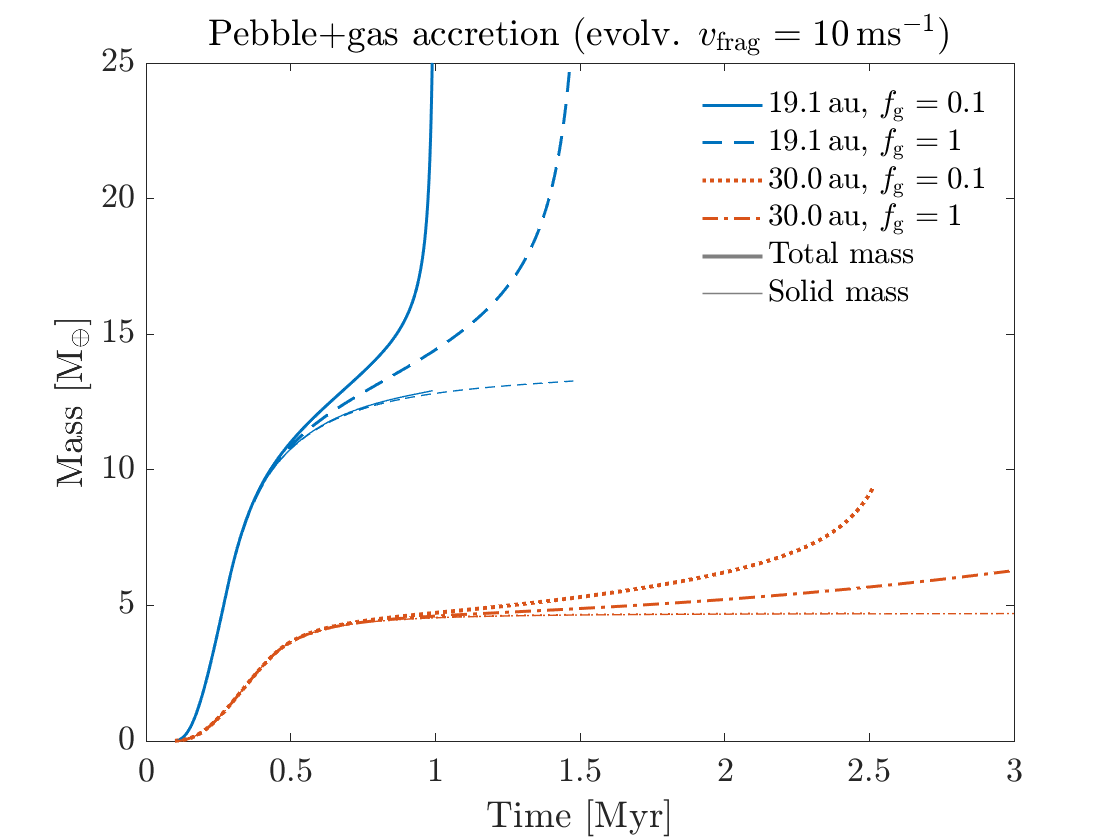}}
{\includegraphics[width=\columnwidth]{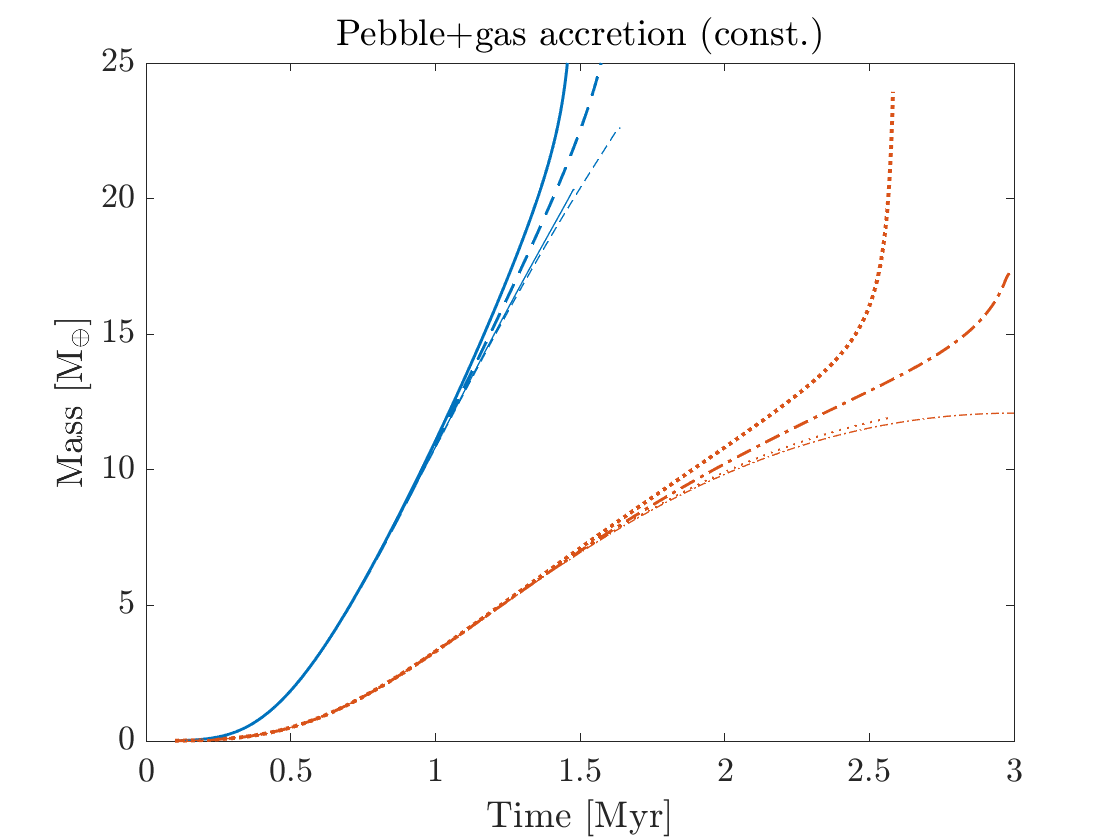}}
{\includegraphics[width=\columnwidth]{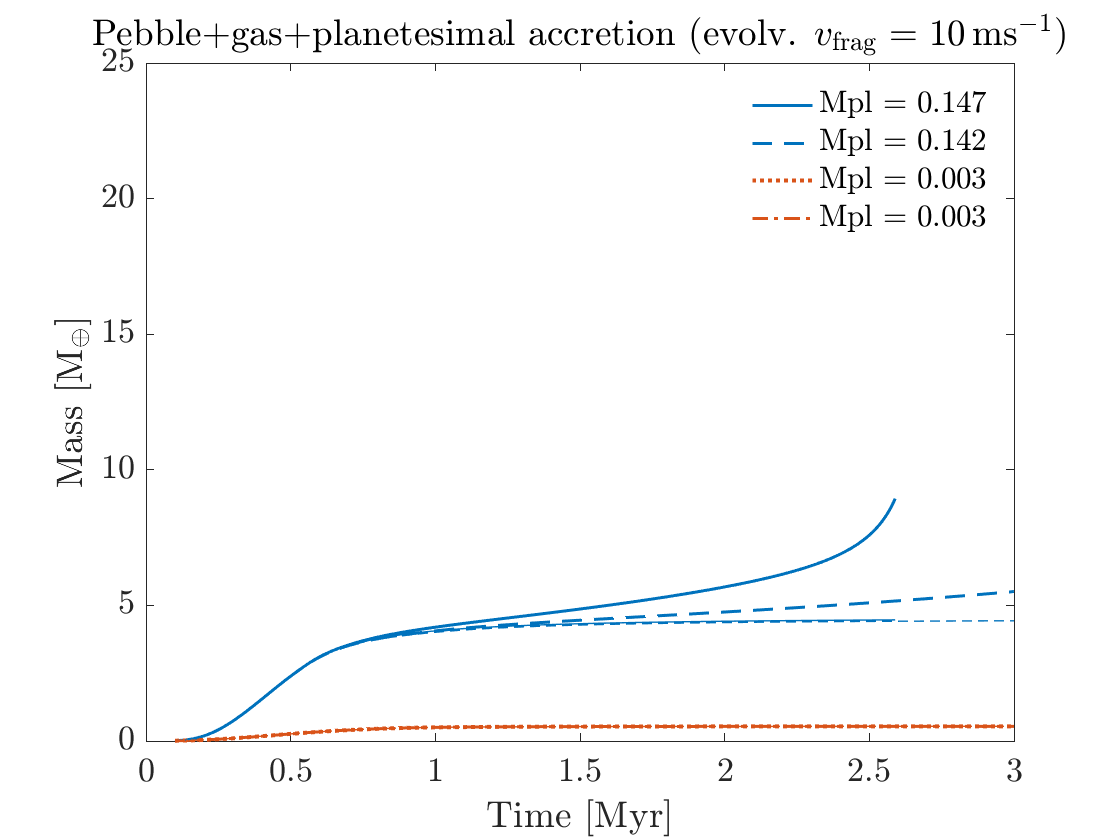}}
{\includegraphics[width=\columnwidth]{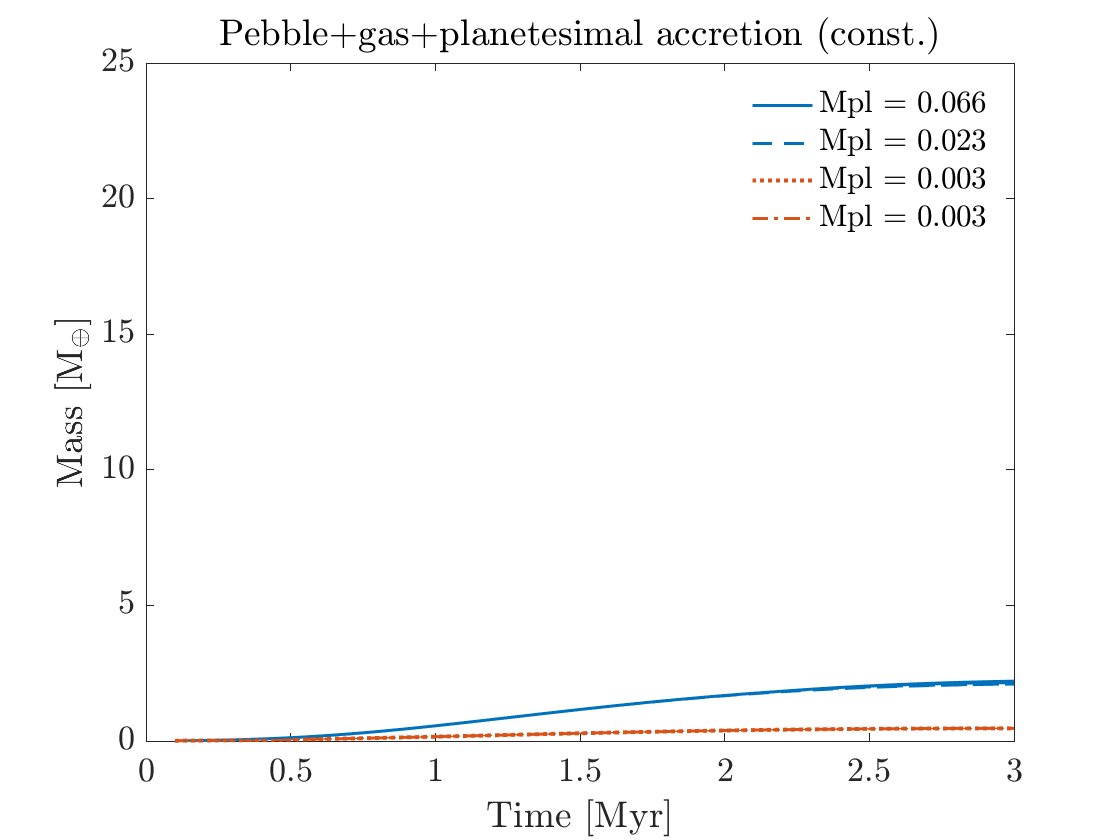}}
\caption{Examples of growth-tracks when the planets form \textit{in-situ}, produced using the parameters given in Table \ref{tab: 2 cases parameters}. The different panels demonstrate how the growth-tracks vary when different accretion mechanisms are considered, and when we switch between the \texttt{constant} and \texttt{evolving} model. }
\label{fig: 2 cases growth track}
\end{figure*}

Fig. \ref{fig: 2 cases growth track} presents the growth-tracks that are obtained when using the simulation parameters specified in Table \ref{tab: 2 cases parameters}. We show how these growth tracks vary when considering growth via only pebble accretion (top row); growth via pebble and gas accretion (middle row); and growth via pebble, gas and planetesimal accretion when assuming that 50\% of the solid mass is locked up in planetesimals (bottom row). Furthermore, we demonstrate how the results change when switching between the \texttt{constant} model (right panels) and the more realistic \texttt{evolving} model (left panels). A comparison between the pebble fluxes and Stokes numbers used in each of the two models is shown in Fig. \ref{fig: 2 cases flux stokes}.  

Let us begin with analysing the case when all growth occurs via pebble accretion, and the pebble flux and Stokes number is obtained from the \texttt{pebble predictor} code (top left panel). Both planets grow several times more massive when we increase the fragmentation velocity from $1\, \textrm{ms}^{-1}$ to $10\, \textrm{ms}^{-1}$. The mass of the outer planet is several times lower than the mass of the inner planet. When we switch to using the \texttt{constant} model with a constant Stokes number and a pebble flux proportional to the disk accretion rate (top right panel), the planets grow more massive. This is mostly because the pebble flux does not decrease as fast with time as it does in the more realistic \texttt{evolving} model. We also show how the growth of the inner planet becomes slower, when we remove the pebbles that are accreted onto the outer planet from the flux towards the inner planet.  

When we take into account the accretion of gas onto the planetary cores (middle panels), the growth-tracks change drastically. The inner planet obtains a H-He mass fraction higher than 50\% and is removed from the simulation, regardless of the pebble model and grain opacity that is being used. The outer planet suffers the same fate when we use the lower grain opacity, despite the core mass being lower than $5\, \textrm{M}_{\oplus}$ in the \texttt{pebble predictor} model. This is a common outcome of models where we use the lower grain opacity. This demonstrates the need for a proper treatment of gas accretion in planet formation simulations during the core accretion phase. 

When we trap half of the solid mass in planetesimals (bottom panels), the planetary growth is less efficient. The total amount of accreted planetesimal mass onto the planets is much less than $1\, \textrm{M}_{\oplus}$. 
The cause for the difference in planetary growth between the case with and without planetesimal accretion is due to the difference in pebble mass. When we trap a significant fraction of the solid mass in planetesimals, the pebble flux decreases and the planetary growth is limited.  

\begin{table}
	\centering
	\caption{Parameters of the example case presented in Fig. \ref{fig: 2 cases flux stokes} and \ref{fig: 2 cases growth track}.}
	\label{tab: 2 cases parameters}
	\begin{tabular}{ll} 
    \hline
    "Representative" St ($Z_{\rm pl}=0$) & 0.021 \\
    "Representative" St ($Z_{\rm pl}=0.5Z$) & 0.015 \\
    $Z$ & 0.02 \\
    $\alpha_{\textrm{T}}$ & $5 \times 10^{-5}$ \\
    $\dot{M}_{0,\textrm{disk}}$ & $6\times 10^{-8}\, \textrm{M}_{\odot}\textrm{yr}^{-1}$ \\
    $t_{\textrm{start}}$ & $10^5\, \textrm{yr}$ \\
    $t_{\textrm{disk}}$ & $3 \times 10^6\, \textrm{yr}$ \\
    $R_{\textrm{edge}}$ (\texttt{evolving}) & $100\, \textrm{au}$ \\
    \hline
	\end{tabular}
\end{table}

\section{Details of the gas accretion model}\label{sect: appendix gas accretion}
To calculate self-consistent the gas accretion rates we use a separate evolution model for the gas accretion \citep[see][for details]{Valletta_2020}. We use MESA version 10108 and the MESA SDK of the same version.\\
Gas accretion rates are calculated in the following way. Every timestep we compute the accretion radius, which is set by:
\begin{align}
    R_{\rm acc} = \frac{GM_{\rm p}}{c_{s}^{2} / k_{1} + G M_{\rm p} / \left ( k_{2} R_{H} \right )},
\end{align}
where G is the universal gravitational constant, M$_{\rm p}$ is the mass of the proto-planet, c$_{s}$ is the sound speed in the disk at the location where the planet is forming and R$_{H}$ is the proto-planet's Hill radius. $k_{1}$ and $k_{2}$ are constants to account for limited supply of gas at the proto-planet's formation location due to disk perturbations. These are set to $\frac{1}{2}$ and $\frac{1}{4}$ respectively, which were found suitable values by \citet{Lissauer_2009}, albeit based on simulations of Jupiter's formation. The values of $k_{1}$ and $k_{2}$ could be close to unity for less massive planets that do not perturb the disk; however, there can be other effects which reduce gas accretion (see Section \ref{sect: gas acc uncertainties}). We then add gas to the envelope until the radius of the proto-planet equals the accretion radius. To construct the fit, we simulate the growth of 30 planets with varying disk and planet parameters, that all have final total masses between 1 and $30\, \textrm{M}_{\oplus}$ if growth occurs via only pebble accretion with the \texttt{constant} model.

The gas accretion rate depends on three things. First, as the planetary mass increases the   accretion radius is extended.  Second, the luminosity of the growing core, which is set by the pebble accretion rate, inflates the envelope. Therefore less gas is needed to fulfil the accretion criterion of R$_{\rm acc} = R_{\rm p}$. Third, the assumed grain opacities can notably limit the cooling efficiency of the accreted gas, which can further enhance the effect that the luminosity has on the gas accretion. Following \citet{Valencia_2013} we combine Rossland opacities from \citet{Freedman_2014} ($\kappa_{mol}$) and grain opacities based on tables from \citet{Alexander_1994} ($\kappa_{grain}$). So that the opacity is:
\begin{equation}
    \kappa_{\textrm{tot}} = \kappa_{\textrm{mol}} + f_{\textrm{g}} \cdot \kappa_{\textrm{grain}}
\end{equation}
We consider cases where $f_{g}=$ 1 and $f_{g}=$ 0.1.

\begin{figure*}
    \includegraphics[width=16cm]{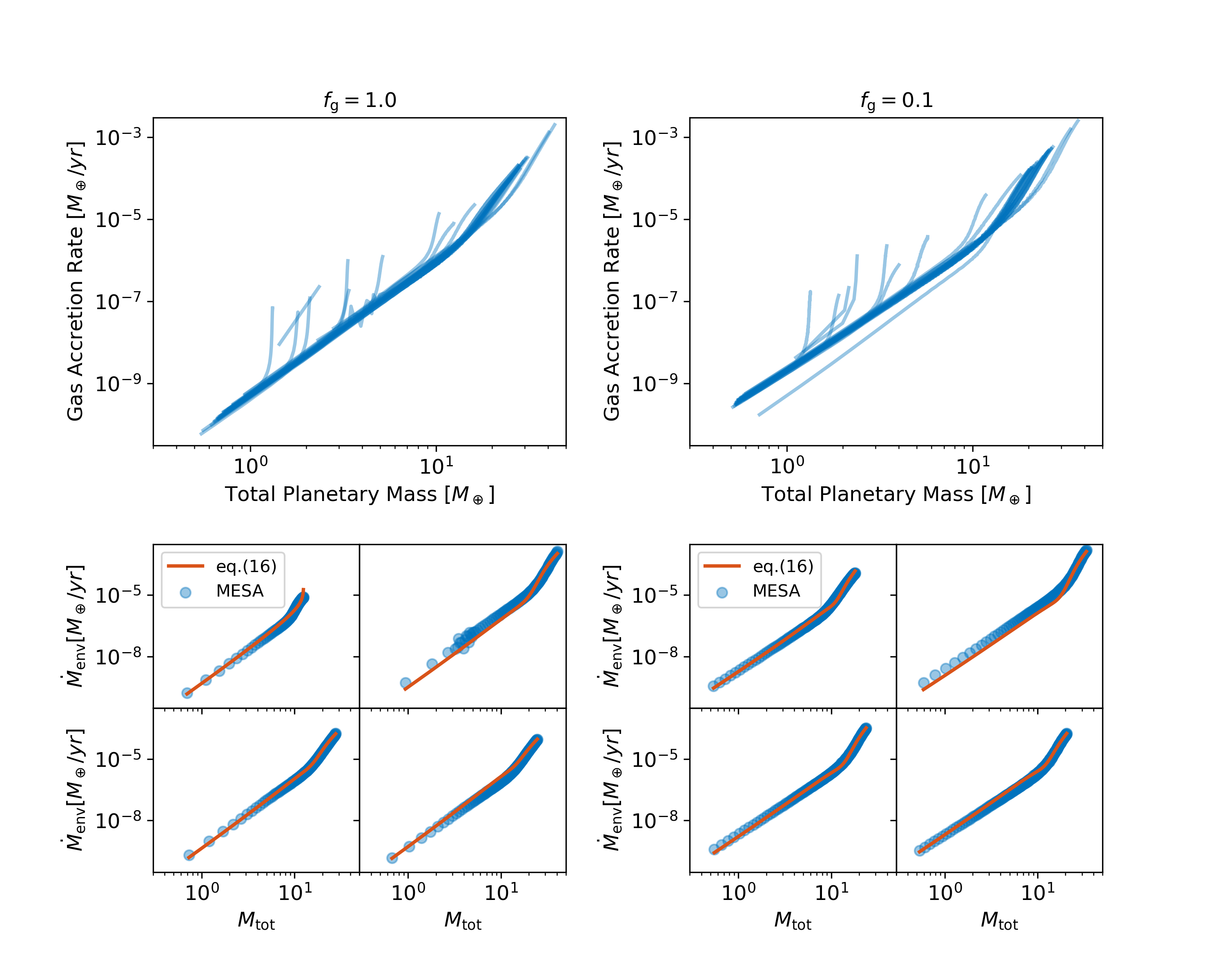}
    \caption{
    {\bf Upper panels}: 
    Gas accretion rates obtained by MESA.
    Each line corresponds to one simulation, pebble accretion was modelled using the \texttt{constant} model.
    Left and right panel show the cases with $f_{\rm g}=1.0$ and $f_{\rm g}=0.1$, respectively.
    {\bf Lower panels}: 
    Comparison of the obtained fitting function with the original data for 4 cases out of the 30 simulations, for each $f_{\rm g}$.
    The solid lines show the fitting function eq.~\ref{eq:dMenvdt_fit}.
    Circle plots are the averaged data of MESA's simulations.
    } 
    \label{fig:Fitting_MESA}
\end{figure*}

Since the original MESA data numerically oscillates in every step, we average the original data for every 100 steps and use the averaged data for fitting functions.
Upper panels in Fig.~\ref{fig:Fitting_MESA} show the averaged gas accretion rate.
When the envelope's mass is small, the gas accretion rate depends on the core mass $M_{\rm core}$.
As the envelope's mass increases, self gravity of the planetary envelope also regulates the gas accretion rate.
From the MESA simulations, we find that the gas accretion rate starts to depend on the envelope mass $M_{\rm env}$ once the envelope-core mass fraction $f_\mathrm{e/c}=M_\mathrm{env}/M_\mathrm{core}$ exceeds $\sim 0.1$.
In order to follow each characteristic of gas accretion, we divide the dataset into two sub-datasets using $f_\mathrm{e/c}$; 
the dataset with $f_\mathrm{e/c}<0.1$ and the dataset with $f_\mathrm{e/c}>0.5$.
The former dataset is used for fitting eq.~\ref{eq:dMenvdt_1}, and the latter one is used for fitting eq.~\ref{eq:dMenvdt_2}, respectively.
We fit the functions to each dataset in logarithm space because the gas accretion rate changes orders of magnitude during the formation of Uranus and Neptune.

The lower panels in Fig.~\ref{fig:Fitting_MESA} show the comparison of MESA and the obtained fitting function eq.~\ref{eq:dMenvdt_fit}.
We do not fit the function when the gas accretion regimes shifts from eq.~\ref{eq:dMenvdt_1} to eq.~\ref{eq:dMenvdt_2} ($0.1 < f_\mathrm{c/e} < 0.5$), but the summation of eq.~\ref{eq:dMenvdt_1} and eq.~\ref{eq:dMenvdt_2} (eq.~\ref{eq:dMenvdt_fit}) can reproduce MESA's results well.
We use eq.~\ref{eq:dMenvdt_fit} in all the simulations used in  this study.

\section{Planetesimal accretion model}\label{sect:appendix planetesimal accretion}
The mean collision rate in eq.~\ref{eq:planetesimal_accretion_rate} can be written as \citep{Inaba2001}: 
\begin{align}
    P_{\rm coll} = {\rm min} \left(P_{\rm med}, \left( {P_{\rm high}}^{-2} +{P_{\rm low}}^{-2} \right)^{-1/2} \right),
\end{align}
where
\begin{align}
    P_{\rm high} &= \frac{{\tilde{r}}^2}{2 \pi}             \left( I_{\rm F} (\beta) +\frac{6 I_{\rm G} (\beta)}{\tilde{r} {\tilde{e}}^2 } \right), \\
    P_{\rm med}  &= \frac{{\tilde{r}}^2}{4 \pi \tilde{i}}   \left( 17.3 +\frac{232}{\tilde{r}} \right), \\
    P_{\rm low}  &= 11.3 {\tilde{r}}^{1/2},
\end{align}
with
\begin{align}
    \tilde{e} &= a_p e/R_{\rm H}, \\
    \tilde{i} &= a_p i/R_{\rm H}, \\
    \tilde{r} &= R_{\rm cap}/R_{\rm H},
\end{align}
where $e$ and $i$ are mean eccentricity and inclination of planetesimals, $a_p$ is the semi-major axis of the protoplanet, $R_{\rm cap}$ is the capture radius of the protoplanet.
$I_{\rm F}$ and $I_{\rm G}$ are given by \citep{Chambers2006}: 
\begin{align}
    I_{\rm F} (\beta) &\simeq \frac{1+0.95925 \beta +0.77251 \beta^2}{\beta (0.13142 +0.12295 \beta)}, \\
    I_{\rm G} (\beta) &\simeq \frac{1 +0.3996 \beta }{\beta (0.0369 +0.048333 \beta +0.006874 \beta^2)},
\end{align}
where $\beta=\tilde{i}/\tilde{e}$.
The evolution of $\tilde{e}$ and $\tilde{i}$ is determined by the viscous stirring between planetesimals, the viscous stirring from the protoplanet, and the gas drag from the gaseous disk.
The rates of the changes in the eccentricity and inclination of planetesimals are given by: 
\begin{align}
    \frac{{\rm d} e^2}{{\rm d} t} &=  \left. \frac{{\rm d} e^2}{{\rm d} t} \right|_{\rm drag} +\left. \frac{{\rm d} e^2}{{\rm d} t} \right|_{\rm VS,M} +\left. \frac{{\rm d} e^2}{{\rm d} t} \right|_{\rm VS,m}, \label{eq:plts_ecc} \\
    \frac{{\rm d} i^2}{{\rm d} t} &=  \left. \frac{{\rm d} i^2}{{\rm d} t} \right|_{\rm drag} +\left. \frac{{\rm d} i^2}{{\rm d} t} \right|_{\rm VS,M} +\left. \frac{{\rm d} i^2}{{\rm d} t} \right|_{\rm VS,m}. \label{eq:plts_inc} 
\end{align}
The gas damping rates are given by \citep{Adachi_1976,Inaba2001}:
\begin{align}
    \frac{{\rm d} e^2}{{\rm d} t} &= - \frac{2 e^2}{\tau_{\rm aero,0}} \left( \frac{9}{4} \eta^2 +\frac{9}{4\pi} \zeta^2 e^2 +\frac{1}{\pi} i^2 \right)^{1/2}, \\
    \frac{{\rm d} i^2}{{\rm d} t} &= - \frac{  i^2}{\tau_{\rm aero,0}} \left(             \eta^2 +\frac{1}{ \pi} \zeta^2 e^2 +\frac{4}{\pi} i^2 \right)^{1/2},
\end{align}
where $\zeta\sim1.211$ and $\tau_{\rm aero,0}$ is given by: 
\begin{align}
    \tau_{\rm aero,0} &=  \frac{2 m_{\rm pl}}{ C_{\rm d} \pi R_{\rm pl}^2 \rho_{\rm gas} v_{\rm K}}, \label{eq:aerodynamic_gas_drag_timescale0}
\end{align}
where $m_{\rm pl}$ is the mass of a planetesimal, $C_{\rm d}$ is the drag coefficient and is set to 1, $R_{\rm pl}$ is the radius of a planetesimal, $\rho_{\rm gas}$ is the density of the disk gas, and $v_{\rm K}$ is the Kepler velocity.
We assume the vertical isothermal disk and use the mid-plane density for $\rho_{\rm gas}$.
The excitation rates of mean square orbital eccentricities and inclinations by the protoplanet are given by \citep{Ohtsuki_2002}: 
\begin{align}
    \left. \frac{{\rm d} e^2}{{\rm d} t} \right|_{\rm VS,M} &= \left( \frac{M_{\rm p}}{3 b M_{*} P_{\rm orb}} \right) P_{\rm VS}, \\
    \left. \frac{{\rm d} i^2}{{\rm d} t} \right|_{\rm VS,M} &= \left( \frac{M_{\rm p}}{3 b M_{*} P_{\rm orb}} \right) Q_{\rm VS},
\end{align}
where $b$ is the full width of the feeding zone and is set to  $10$, $M_{*}$ is the central star's mass, and $P_{\rm VS}$ and $Q_{\rm VS}$ are given by: 
\begin{align}
    P_{\rm VS} &= \frac{73 {\tilde{e}}^2}{10 \Lambda^2} \ln \left(1 +10\frac{\Lambda^2}{{\tilde{e}}^2} \right)
                    + \frac{72 I_{\rm PVS} (\beta)}{\pi \tilde{e}\tilde{i}} \ln \left(1 +\Lambda^2 \right), \\
    Q_{\rm VS} &= \frac{4 {\tilde{i}}^2 +0.2 \tilde{i}{\tilde{e}}^3 }{10 \Lambda^2 \tilde{e}} \ln \left(1 +10 \Lambda^2 {\tilde{e}} \right)
                    + \frac{72 I_{\rm QVS} (\beta)}{\pi \tilde{e}\tilde{i}} \ln \left(1 +\Lambda^2 \right),
\end{align}
where $\Lambda=\tilde{i} ({\tilde{e}}^2+{\tilde{i}}^2)/12$.
For $0 < \beta \leq 1$, $I_{\rm PVS}$ and $I_{\rm QVS}$ can be approximated by \citep{Chambers2006}: 
\begin{align}
    I_{\rm PVS} (\beta) \simeq \frac{\beta -0.36251}{0.061547 +0.16112 \beta +0.054473 \beta^2}, \\
    I_{\rm QVS} (\beta) \simeq \frac{0.71946 -\beta}{0.21239  +0.49764 \beta +0.14369  \beta^2}.
\end{align}
We also consider the excitation rates of mean square orbital eccentricities and inclinations by the mutual interactions of planetesimals are given by \citep{Ohtsuki_2002}: 
\begin{align}
    \left. \frac{{\rm d} e^2}{{\rm d} t} \right|_{\rm VS,m} &= \frac{1}{6} \sqrt{\frac{G a_p}{M_{*}} } \Sigma_\mathrm{pl} h_\mathrm{m} P_{\rm VS}, \\
    \left. \frac{{\rm d} i^2}{{\rm d} t} \right|_{\rm VS,m} &= \frac{1}{6} \sqrt{\frac{G a_p}{M_{*}} } \Sigma_\mathrm{pl} h_\mathrm{m} Q_{\rm VS},
\end{align}
with
\begin{align}
    h_\mathrm{m} = \left( \frac{2 m_\mathrm{pl}}{3 M_{*}} \right)^{1/3}.
\end{align}.

\section{Successful analogues when the allowed mass difference is doubled}\label{sect: double mass difference}
\begin{figure*}
\centering
{\includegraphics[width=\columnwidth]{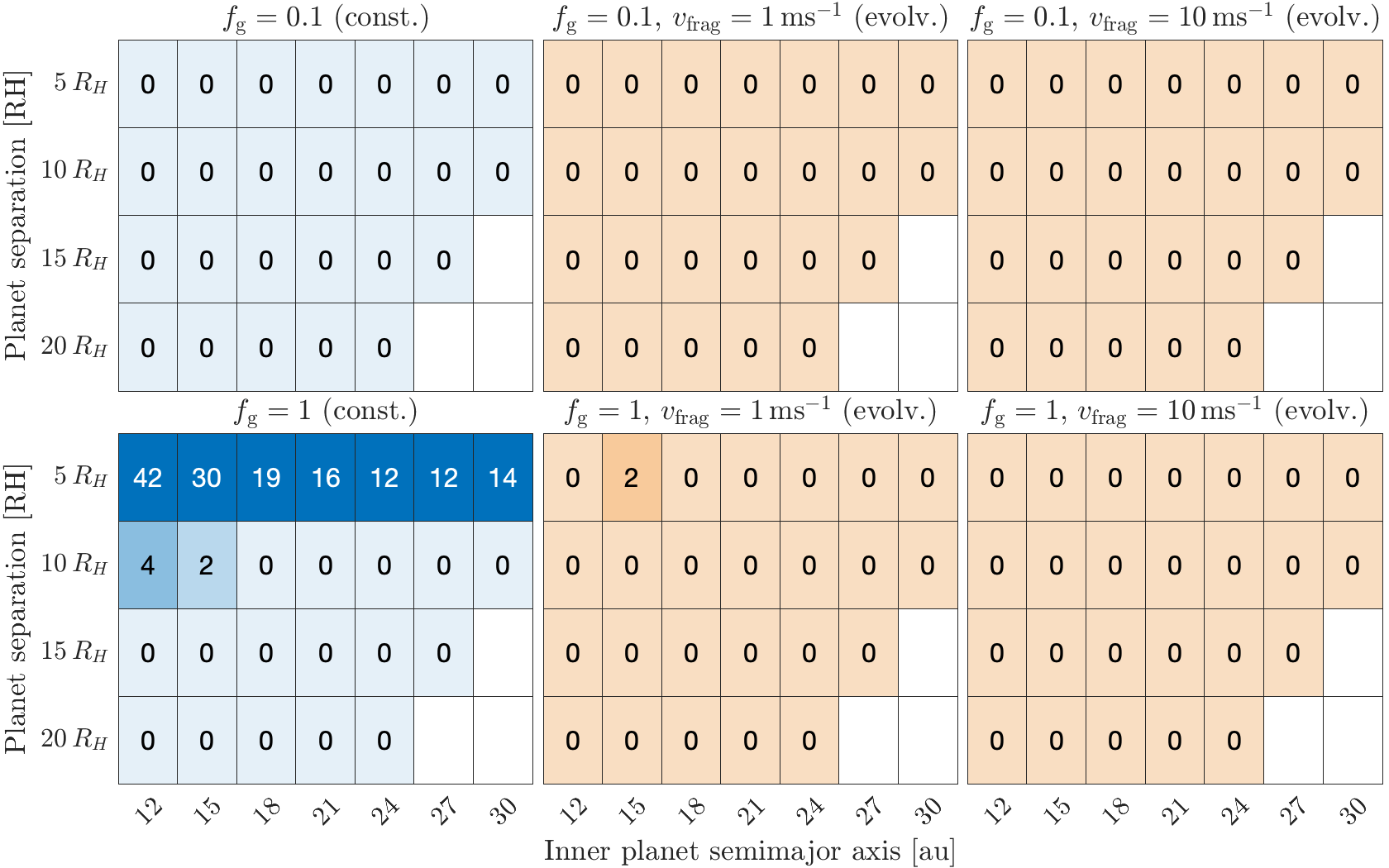}}
{\includegraphics[width=\columnwidth]{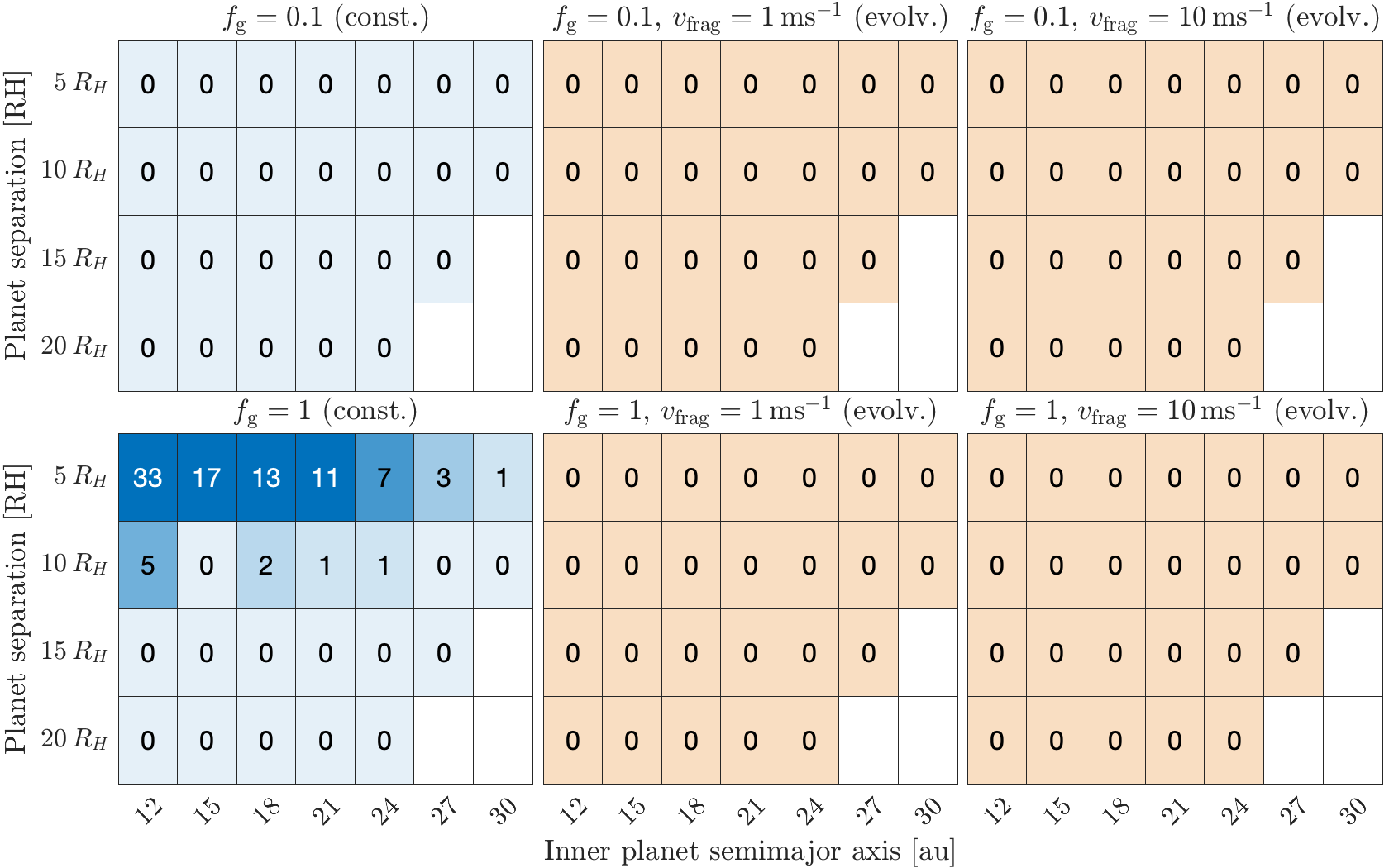}}
{\includegraphics[width=\columnwidth]{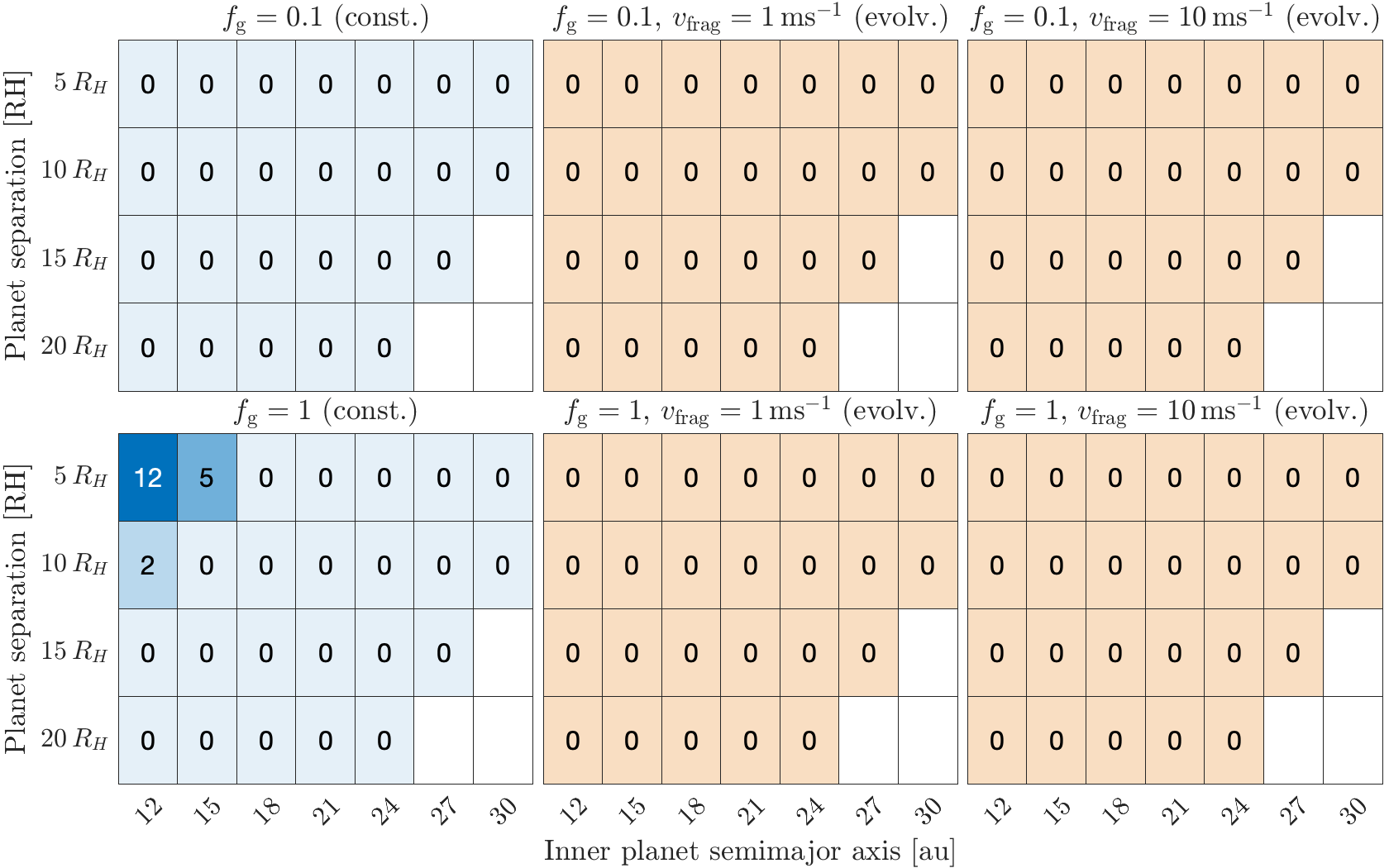}}
\caption{Same as Fig. \ref{fig: success pebbGasPlan}, but the allowed mass difference was increased to $3\, \textrm{M}_{\oplus}$, resulting in more successful analogues. }
\label{fig: success pebbGasPlan 3Me}
\end{figure*}

\begin{figure*}
\centering
{\includegraphics[width=\columnwidth]{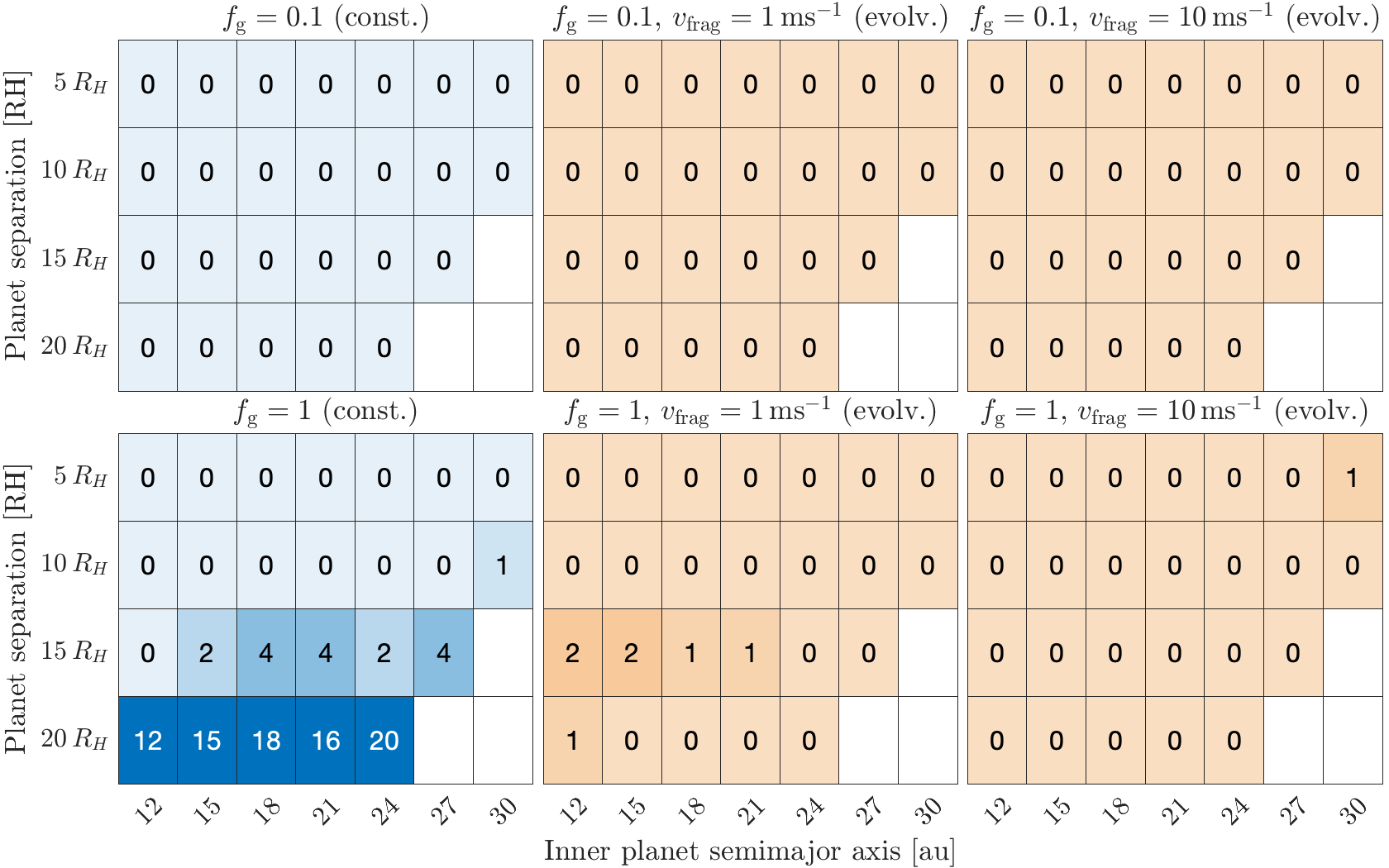}}
{\includegraphics[width=\columnwidth]{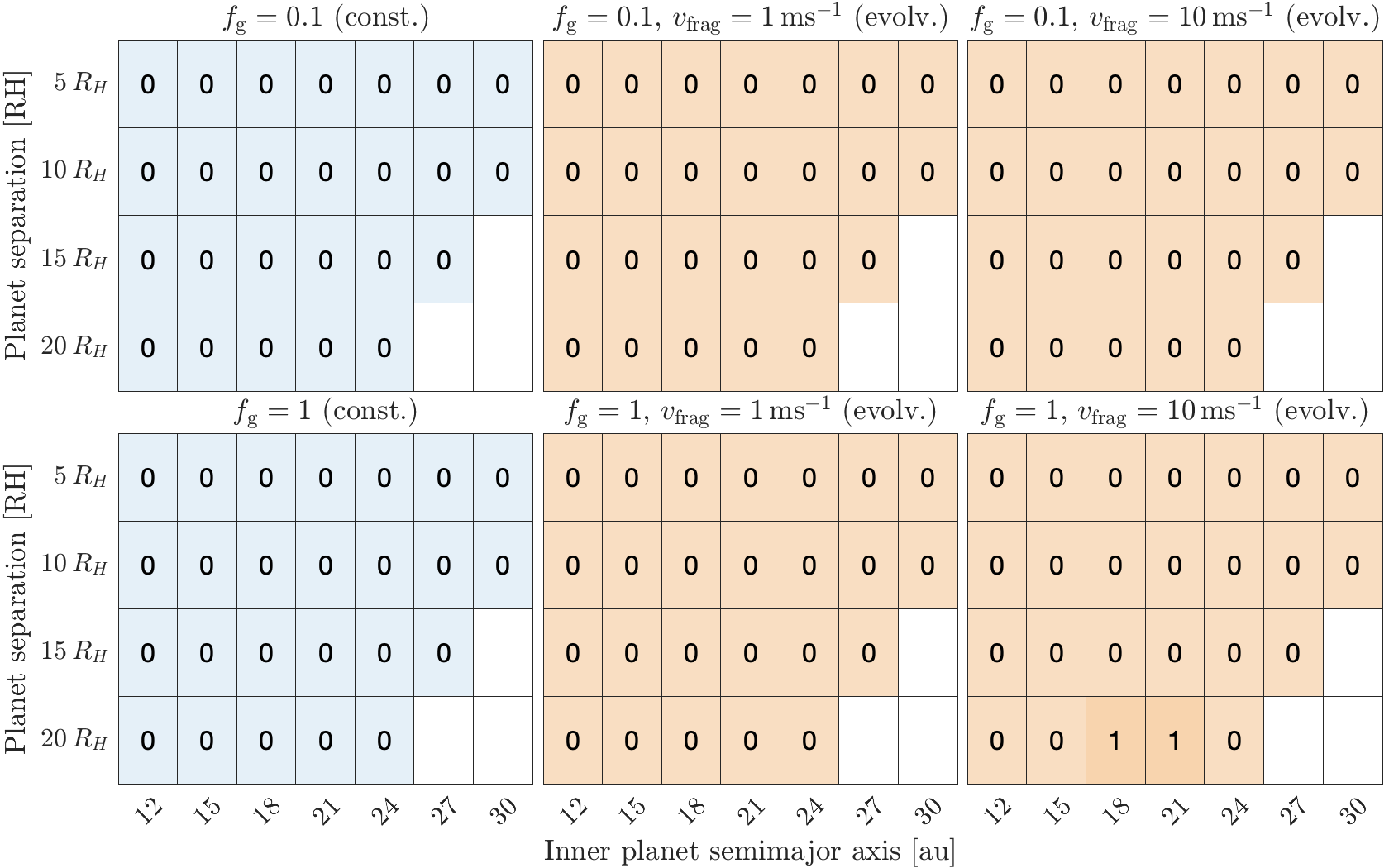}}
\caption{Same as Fig. \ref{fig: success pebbGas nLTu}, but the allowed mass difference was increased to $3\, \textrm{M}_{\oplus}$, resulting in more successful analogues.}
\label{fig: success pebbGas nLTu 3Me}
\end{figure*}

In the main text, we identified successful Uranus and Neptune analogues by finding all simulations where the total mass of the inner and outer planet at the time of disk dissipation was within $1.5\, \textrm{M}_{\oplus}$ of $14.5\, \textrm{M}_{\oplus}$ and $17.1\, \textrm{M}_{\oplus}$, respectively. Then we removed all simulations where the H-He mass fraction of any of the planets was above 20\% at disk dissipation. Furthermore, we also allowed for the planets to switch places after formation. In other words, we also searched for simulations where the mass of the inner planet was within $1.5\, \textrm{M}_{\oplus}$ of the current mass of Neptune, and the mass of the outer planet was within $1.5\, \textrm{M}_{\oplus}$ of the current mass of Uranus. Figure \ref{fig: success pebbGasPlan 3Me} and \ref{fig: success pebbGas nLTu 3Me} show the number of successful analogues that was obtained when we increased the allowed mass difference to $3\, \textrm{M}_{\oplus}$. 

In Figure \ref{fig: success pebbGasPlan 3Me} where the planetary embryos form and begin to accrete mass at the same time, increasing the allowed mass difference results in many more successful analogues while using the \texttt{constant} model. However, when using the more realistic \texttt{evolving} model, there is almost no change. There are two successful analogues found in the case with no planetesimal accretion, and none when there is planetesimal accretion. 
When the outer embryo is formed with a higher mass (Fig. \ref{fig: success pebbGas nLTu 3Me}), increasing the allowed mass difference more than doubles the amount of successful analogues found in both the \texttt{constant} model and the \texttt{evolving} model. 

\bsp	
\label{lastpage}
\end{document}